\documentclass[aps, prd, twocolumn, lengthcheck, superscriptaddress, longbibliography, nofootinbib]{revtex4-2}

\usepackage{times}
\usepackage{amsmath,amssymb,amsfonts}
\usepackage{bm}
\usepackage{graphicx}
\DeclareGraphicsExtensions{.pdf,.eps,.png,.jpg}
\usepackage{acronym}
\usepackage{color}
\usepackage{xcolor}
\usepackage[caption=false]{subfig}
\usepackage{tabularx}
\usepackage{longtable}
\usepackage{multirow}
\usepackage{dcolumn}
\usepackage{soul}
\usepackage{float}
\usepackage{hyperref}
\usepackage{cleveref}
\usepackage{makecell}

\graphicspath{{figures/}}

\usepackage[normalem]{ulem}

\begin{document}

\title{Can we discern millilensed gravitational-wave signals from signals produced by precessing binary black holes with ground-based detectors?}

\author{Anna Liu}
\email{ania.liu@link.cuhk.edu.hk}
\affiliation{Department of Physics, The Chinese University of Hong Kong, Shatin, New Territories, Hong Kong}

\author{Kyungmin Kim}
\email{kkim@kasi.re.kr}
\thanks{Corresponding author}
\affiliation{Korea Astronomy and Space Science Institute, 776 Daedeok-daero, Yuseong-gu, Daejeon 34055, Republic of Korea}

\date{\today}

\begin{abstract}
Millilensed gravitational waves (GWs) can potentially be identified by the interference signatures caused by $\sim\!\mathcal{O}(10\textrm{--}100)~\textrm{ms}$ time delays between multiple overlapping lensed signals. 
However, distinguishing millilensed GWs from GWs generated by precessing binary black-hole mergers can be challenging due to their apparent similar waveform shapes.
This morphological similarity may be an obstacle to template-based searches to correctly identifying the origin of observed GWs and poses a fundamental question: \emph{Can we discern millilensed GW signals from signals produced by precessing binary black holes?} 
In this study, we investigate the feasibility of distinguishing between these GWs by performing a proof-of-principle injection study of simulated millilensed precessing GW signals, within the context of ground-based LIGO-Virgo-KAGRA detector network detections. 
We compare signal-to-noise ratios (SNRs) computed using templates with different hypotheses (lensed or unlensed and precessing or nonprecessing) for the target signal.
We further show from the parameter estimation study that while the recovery of lensing magnification is sensitive to the inclusion of precession in the recovery model, it is possible to identify millilensing in precessing GW signals with an SNR of 18.
The recovery of precession in the presence of lensing is more challenging but improves significantly for signals with an SNR of 40.
Nonetheless, neglecting millilensing effects results in biases in the recovered spins, revealing the importance of accounting for these effects in accurate GW signal analysis.
\end{abstract}

\maketitle

\section{Introduction}

Gravitational waves (GWs), predicted by Einstein's theory of general relativity (GR), have significantly enriched our understanding of the Universe, providing invaluable insights into coalescing compact objects and the fundamental nature of gravity~\cite{abbott2016observation, abbott2019gwtc1, abbott2021gwtc2, abbott2021gwtc3}. 
Further exploration in GW astrophysics can unveil previously inaccessible phenomena, including gravitational lensing of GWs induced by massive objects.
Analogous to electromagnetic radiation, GWs can be subject to gravitational lensing effects due to intervening massive objects between the source and the observer~\cite{Ohanian:1974, Bliokh:1975apss, Bontz:1981apss, Thorne:1983, Deguchi:1986apj, Schneider:1992, Nakamura:1999ptps, Takahashi:2003ix}.  
This phenomenon introduces observable modifications in the GW signal, including amplitude magnification, phase shift and delayed arrival time relative to a nonlensed GW~\cite{Takahashi:2003ix, cao2014gravitational, Dai2017OnWaves, Ezquiaga_2021}. 
Although searches for GW lensing have been conducted in the current GW data, no confirmed cases have been found~\cite{Hannuksela:2019kle, abbott2021search, abbott2023search, Janquart_2023}. 
A possible obstacle in finding lensed signatures may lie in their morphological similarity with other physical effects, including spin-induced precession, explored in this paper.

In scenarios when short-duration (transient) GW signals emitted by coalescing compact binaries encounter sufficiently small lenses (with masses $M_{Lz}\lesssim\!10^5 M_\odot$), multiple lensed GWs overlap, yielding a single superposed GW at the detector, a phenomenon commonly referred to as microlensing and millilensing~\cite{Diego:2019lcd, Seo_microlensing, Kim:2020xkm, Kim:2022lex,
Lai2018DiscoveringLensing, Dai2018DetectingWaves, Liao2018AnomaliesSubstructures, Christian:3Gdetections, Pagano2020LensingGW:Waves, Wright2021Gravelamps:Selection, Liu2023exploring}~\footnote{Some literature, including~\cite{abbott2021search, abbott2023search, Kim:2020xkm, Kim:2022lex}, collectively refers to lensing by lens masses $M_{Lz}\lesssim\!10^5 M_\odot$ as microlensing, without distinguishing the geometric-optics millilensing regime considered in this work.}.
Throughout this work, we collectively refer to \textit{millilensing} as lensing in the geometric optics regime with time delays between multiple lensed GWs of the order 
of $\sim\!\mathcal{O}(10\textrm{--}100)~\textrm{ms}$ and corresponding lens masses in the range of $10^2-10^5 M_\odot$ for ground-based GW detectors~\cite{Liu2023exploring}.
In this framework, we do not consider microlensing, which would account for even smaller lens masses ($M_{Lz}\lesssim\!10^2 M_\odot$) and involve wave optics effects. 
With the time delays between multiple lensed GWs shorter than the duration of signals from compact binary coalescences, individual lensed GWs overlap, leading to a frequency-dependent interference pattern, commonly referred to as a \emph{beating pattern}, a characteristic signature of GW millilensing. 
By identifying beating patterns in GW signals, we can potentially classify them as millilensed GWs.

\begin{figure*}[t!]
\includegraphics[width=1.\linewidth]{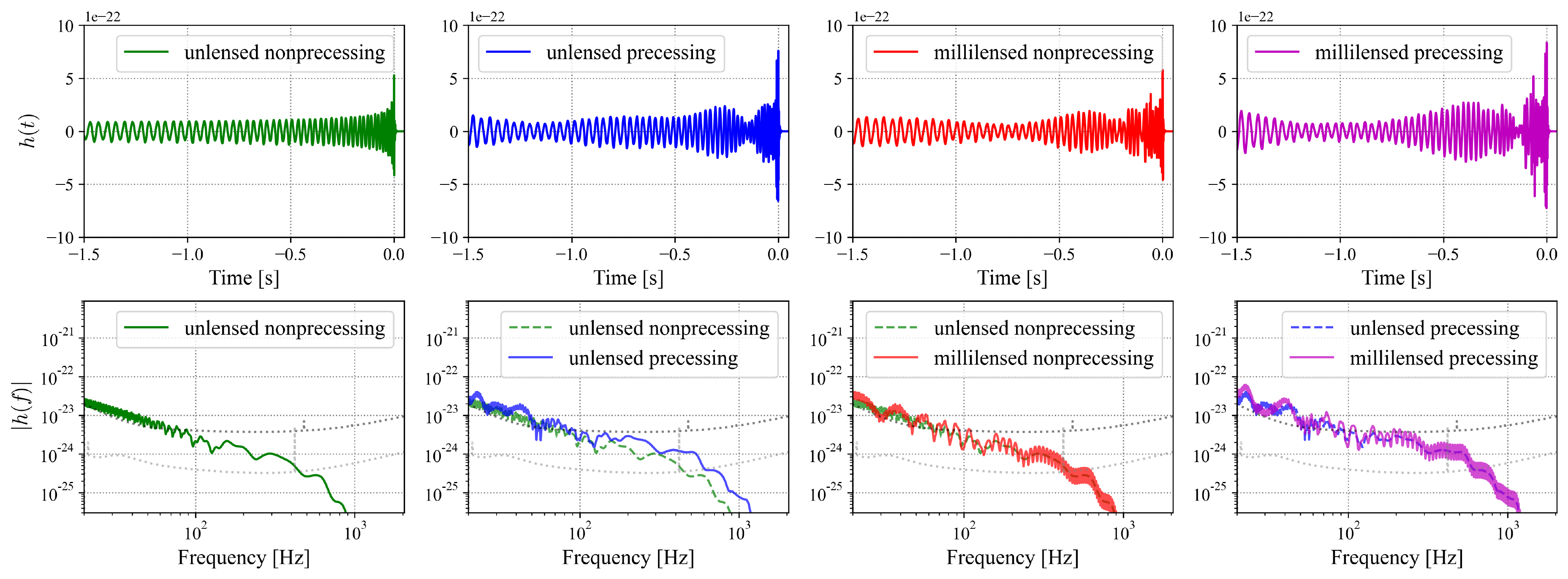}
\caption{Noise-free signals studied in this paper. The top and bottom panels show $h(t)$ and $|h(f)|$, respectively. From the left to the right, we present unlensed nonprecessing, unlensed precessing, millilensed nonprecessing, and millilensed precessing signals, respectively. The time of $h(t)$ is given w.r.t. the merging time at $t=0~\textrm{s}$. In $|h(f)|$ plots, we display a counter-example signal (dashed line) and the design sensitivity curves of the Advanced LIGO (aLIGO)~\cite{aligodesignpsd} (dotted dark grey) and the Einstein Telescope (ET)~\cite{Hild:2010id} (dotted light grey) for comparison.}
\label{fig:wf_noise_free}
\end{figure*}

On the other hand, a similar pattern in the GW waveform can occur from precessing binaries. The spin-induced precessional motion of a compact binary system--caused by the misalignment between the spins of the binary components and the system's orbital angular momentum--can introduce GW amplitude modulations, frequency oscillations and a phase shift relative to a nonprecessing GW~\cite{Apostolatos1994spin, Kidder1995, Schmidt2011:trackingsignal, Hannam_2014, Vitale_2014, o2014parameter,khan2019phenomenological, romero2023eccentricity}. 
As shown in Fig.~\ref{fig:wf_noise_free}, the modulation induced by precession during the inspiral (corresponding to $t\!<\!0~\textrm{s}$), appears strikingly similar to the beating pattern exhibited by a millilensed GW signal.

While the observed waveform may appear similar, it is crucial to recognize that the physical origins and applications of these two effects are different. 
Gravitational lensing of light has proven to be a valuable tool for studying the distribution of matter in the Universe. 
Similarly, GW lensing can be used to investigate dark matter substructures, probe high-redshift cosmology, localize black-hole mergers or test predictions of GR~\cite{Takahashi2005AmplitudeUniverse, Itoh2009AWaves,
Baker2016Multi-MessengerWaves,collett2017testing, Liao2017PrecisionSignals, Fan2017SpeedSignals, Lai2018DiscoveringLensing, Dai2018DetectingWaves, Mukherjee2019Multi-messengerWaves, mukherjee2020probing, Diego2019ConstrainingFrequencies,Oguri2020ProbingWaves, Goyal2020TestingSignals, Hannuksela2020LocalizingLensing, Finke2021ProbingBinaries,Iacovelli2022ModifiedFunction, Chung2021LensingMass, Sereno2010StrongLISA, Bolejko2012Anti-lensing:Voids,hernandez2022measuring, Tambalo2022GravitationalMatter,Basak2022ConstraintsMicrolensing, Basak2022ProspectsHole}. 
On the other hand, spin-induced precession of binary systems is directly related to the properties of the binary itself. 
It can provide insights into the formation and evolution history of binary black holes (BBHs), shedding light on the origin of stellar-mass black-hole systems~\cite{Apostolatos1994spin, Kidder1995, rodriguez2016illuminating, harry2016searching,  khan2019phenomenological, kalogera2000spin, mapelli2020formation, johnsonmcdaniel2023distinguishing}.
Therefore, correctly distinguishing between these two effects is of great importance not only in the context of current detections but also for future observations with next-generation GW detectors.

The morphological similarity between the two types of GWs can pose a challenge for template-based analyses, as they require accurate assumptions on the intrinsic and extrinsic parameters of GWs~\cite{cutler1994gravitational, veitch2015parameter}. 
Consequently, the fundamental question arises: \emph{Is it possible to distinguish millilensed GW signals from signals originating from precessing binary black-hole mergers?} 
To answer this question in the context of LIGO-Virgo-KAGRA (LVK)~\cite{Acernese_2008, Abbott_2009, aasi2015advanced, kagra2019kagra} detections, we study simulated precessing and nonprecessing GWs with and without millilensing and assess the possibility of distinguishing the two by (a) comparing the signal-to-noise ratios (SNRs) obtained with different templates and (b) performing a parameter estimation (PE) study to infer the properties of the binary system with simulated example signals of interest. 

By comparing SNRs of example signals, we find that it is likely possible to discriminate between millilensed GWs and GWs from precessing stellar-mass BBHs. However, this conclusion is valid under two conditions: (i) the template is built with the correct intrinsic and extrinsic parameters of the BBH, regardless of the presence of millilensing, and (ii) a proper unlensed or millilensed hypothesis is considered for the template.
Next, we perform a PE study to investigate the possibility of differentiating millilensed GWs from precessing ones. We also aim to address the degeneracies that may arise between the two effects, especially if a precessing signal is millilensed.
From the PE study, we find that the recovery of lensing magnification is sensitive to the inclusion of precession in the recovery model, while time delays are correctly recovered even when neglecting precession for signals with SNR 18 and higher.
Precessing spins can be accurately recovered for lensed precessing signals with SNR 40 when accounting for lensing. 
However, neglecting the lensing effect leads to biases in the recovered spins. 

\begin{figure*}[t!]
    \includegraphics[width=1.\linewidth]{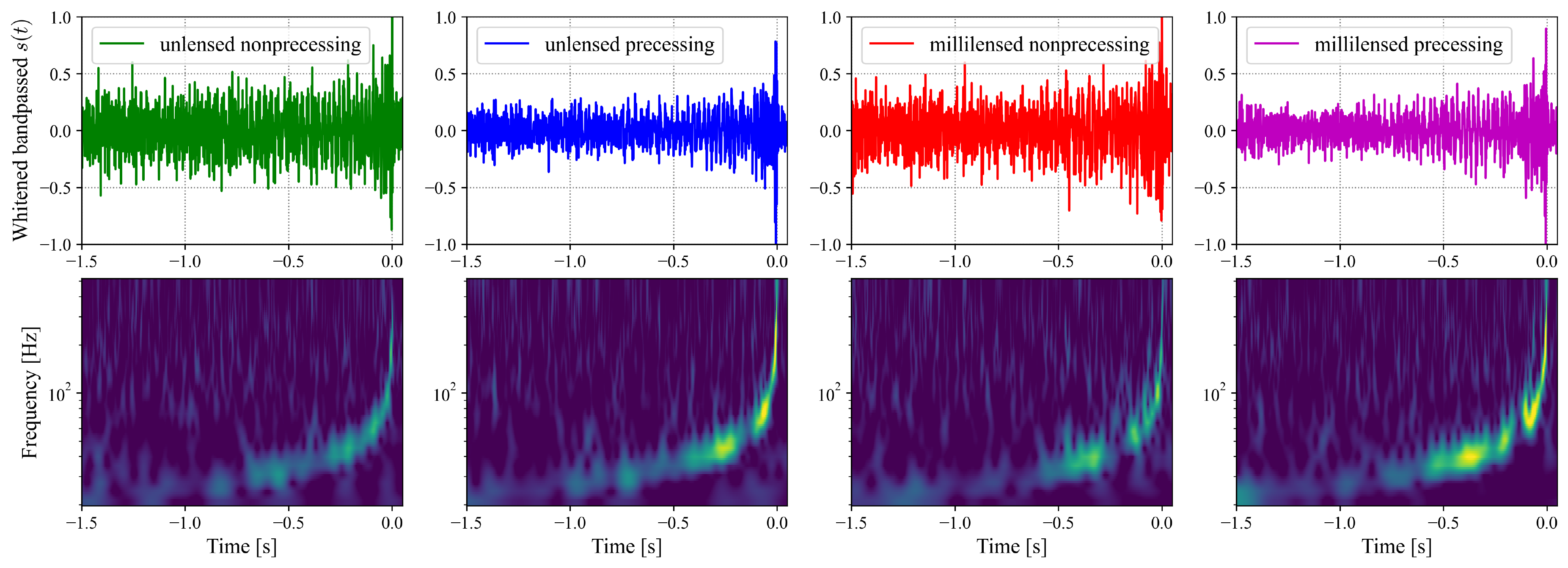} \label{fig:wf_noisy_aligo}
\caption{The same signals of Fig.~\ref{fig:wf_noise_free} but injected into the noise data drawn from the aLIGO's design PSD~\cite{aligodesignpsd}. We plot whitened bandpassed signals $s(t)$ and normalize it with the peak amplitude at $t=0~\textrm{s}$. The bottom panels show the spectrograms of each signal.\label{fig:wf_noisy_aligo}}
\end{figure*}

The paper is organized as follows. 
In the following Sec.~\ref{sec:test_examples}, we first outline the procedure we follow to generate the example signals shown in Fig.~\ref{fig:wf_noise_free} (Sec.~\ref{sec:simulate_GWs}). 
We categorize each signal as either millilensed or precessing, or possessing characteristics of both millilensing and precession (Sec.~\ref{sec:signals_in_noise}).
We compare the computed SNRs of the example signals buried in noise in Sec.~\ref{sec:SNR_comparison} and perform a PE study on the above-generated signals in Section~\ref{sec:PE_example}.
Then we conduct a broader PE study of simulated millilensed and precessing GWs across a wide parameter space to probe which region of the parameter space is most sensitive to the degeneracy between precession and millilensing (Section~\ref{sec:PE_broad_study}). 
Finally, in Section~\ref{sec:conclusions}, we discuss our findings and possible future steps.


\section{Testing distinguishability with example signals}
\label{sec:test_examples}

\subsection{Simulating example GW signals of interest}
\label{sec:simulate_GWs}
For the preparation of signals of interest for the first part of our study, we utilize the \textsc{PyCBC}~\cite{DalCanton:2014hxh,Usman:2015kfa,alex_nitz_2020_4075326} software package and choose the phenomenological waveform models, \textsc{IMRPhenomXPHM} and \textsc{IMRPhenomXHM}~\cite{Garcia-Quiros:2020qpx, Pratten:2020ceb}, not only to incorporate multipole harmonic modes of GWs but to generate both precessing and nonprecessing signals in a consistent manner.

First, we generate unlensed nonprecessing and unlensed precessing signals ($h_\textrm{N}$  and $h_\textrm{P}$, respectively) with the parameters summarized in Table~\ref{tab:params}: 
\begin{table}[t!]
    \caption{Parameters used for the generation of example signals of interest.}
    \centering
    \begin{tabularx}{1.\linewidth}{@{} l l @{\extracolsep{\fill}} c @{}}
    \toprule
         Case & Parameter & Value \\
         \hline
         \multirow{4}{*}{Common} & Mass 1, $m_1$ & $30M_\odot$ \\
         & Mass 2, $m_2$ & $10M_\odot$ \\
         & Luminosity distance, $d_L$ & $500~\textrm{Mpc}$ \\
         & Sampling rate & $8192~\textrm{Hz}$ \\
         \hline
         \multirow{3}{*}{Nonprecessing} & Spin 1, $\Vec{s}_1\!=\!(s_{1x}, s_{1y}, s_{1z})$ & (0, 0, 0) \\
         & Spin 2, $\Vec{s}_2\!=\!(s_{2x}, s_{2y}, s_{2z})$ & (0, 0, 0) \\
         & Inclination angle, $\theta_{LN}\sphericalangle{\Vec{L}\Vec{N}}$ & $80~\textrm{deg}$ \\
         \hline
         \multirow{3}{*}{Precessing} & Spin 1, $\Vec{s}_1\!=\!(s_{1x}, s_{1y}, s_{1z})$ & (0.76, 0.06, 0.55) \\
         & Spin 2, $\Vec{s}_2\!=\!(s_{2x}, s_{2y}, s_{2z})$ & (-0.19, 0.74, -0.33) \\
         & Inclination angle, $\theta_{JN}\sphericalangle{\Vec{J} \Vec{N}}$ & $160~\textrm{deg}$ \\
         \hline
         \multirow{3}{*}{Millilensing} & Source position, $y$ & 0.9 \\
         & Log-redshifted lens mass, & \multirow{2}{*}{$3.2$} \\
         & $\log_{10}\!(M_{lz}/M_\odot)$ & \\
    \hline
    \hline
    \end{tabularx}
    \label{tab:params}
\end{table}
For the nonprecessing signal, we consider a BBH system with nonspinning BHs, i.e., $\Vec{s}_1=\Vec{s}_2=(0,0,0)$. Meanwhile, the spin parameters of the precessing signal are randomly chosen to satisfy the spin magnitude $|\Vec{s}|\!\equiv\!s\!=\!(s_x^2\!+\!s_y^2\!+\!s_z^2)^{1/2}\!\leq\!1$. 
With this choice, the effective precession spin parameter $\chi_p$~\cite{Schmidt:2014iyl} of the precessing binary is calculated to be $\chi_p\simeq\!0.76$ via the \textsc{chi\_p} function of \textsc{PyCBC} with the given component masses and spins $m_1$, $m_2$, $s_{1x}$, $s_{1y}$, $s_{2x}$, and $s_{2y}$.

Next, following~\cite{Takahashi:2003ix}, we simulate a millilensed signal $\tilde{h}_\textrm{ML}$ via $\tilde{h}_\textrm{ML}(f)\!=\!F(f)\tilde{h}_\textrm{U}(f)$ adopting the thin-lens approximation, where $\tilde{h}_\textrm{U}(f)$ denotes an unlensed signal, either $\tilde{h}_\textrm{N}$ or $\tilde{h}_\textrm{P}$, in the frequency domain and $F(f)$ is the amplification factor given as a function of frequency. To specify $F(f)$ enabling us to simulate a $\tilde{h}_\textrm{ML}(f)$, we take the point-mass lens model in the geometrical optics limit\footnote{The sensitive frequency band of current ground-based detectors corresponds to where $|F(f)|$ in the wave optics regime asymptotically converges to the geometrical optics limit for the lens mass of $\lesssim\!10^5M_\odot$.~\cite{Takahashi:2003ix, Kim:2022lex}} which allows $F(f)$ to be written as $F(f)\!=\!\sqrt{|\mu_{+}|}\!-\!i \sqrt{|\mu_{-}|} e^{2\pi i f \Delta t}$, where $\mu_\pm\!=\!1/2\!\pm\!(y^2\!+\!2)/(2y\gamma)$ are the magnification factors for each of the two lensed signals and $\Delta t\!=\!4 G M_{lz} c^{-3} [(y \gamma)/2\!+\!\ln \{ (\gamma\!+\!y)/(\gamma\!-\!y) \} ]$ is the time delay between them, where $y$ is the dimensionless source position parameter, $M_{lz}\!=\!M_l(1+z)$ is the redshifted lens mass, and $\gamma=\sqrt{y^2 + 4}$.

We then convert $\tilde{h}(f)$ to $h(t)$ via the inverse Fourier transformation. We choose $y=0.9$ and $M_{lz}=10^{3.2}M_\odot$ in order to produce an apparently similar $h_\textrm{ML}(t)$ to given $h_\textrm{P}(t)$; the parameters give $\mu_+\!\simeq\!1.21$, $\mu_{-}\!\simeq\!-0.21$, and $\Delta t\!\simeq\!58~\textrm{ms}$.

Finally, to consider realistic detection scenarios, we generate the strain data $s(t)$ such as $s(t)\!=\!h(t)\!+\!n(t)$, where $n(t)$ is a noise data. 
For $n(t)$, we implement the Advanced LIGO (aLIGO)'s design power spectral density (PSD)~\cite{aligodesignpsd}. 
In addition, we employ bandpass filtering for whitened $s(t)$ following the conventional signal processing for the LIGO-Virgo-KAGRA data~\cite{LIGOScientific:2019hgc}, except the frequency window for the bandpass filter: For the purpose of visualization, we apply an empirically determined optimal frequency window $[20~\textrm{Hz}, 400~\textrm{Hz}]$\footnote{Conventionally adopted window for the bandpass filter for the LIGO-Virgo-KAGRA data analysis is [35 Hz, 350 Hz].~\cite{LIGOScientific:2019hgc}}. We present the resultant $s(t)$ in Fig.~\ref{fig:wf_noisy_aligo}. 


\subsection{Comparing apparent morphology between signals in noise}
\label{sec:signals_in_noise}
We examine the signals of interest buried in noise. Hereafter, `NU', `PU', `NL', and `PL' stand for the nonprecessing unlensed, precessing unlensed, nonprecessing (milli)lensed, and precessing (milli)lensed signals, respectively.

\begin{table*}[t!]
    \caption{Matched-filter SNRs of the examined pairs. We can see SNRs of homogeneous pairs are higher than heterogeneous pairs as expected and SNRs of precessing targets are slightly higher than nonprecessing targets. For comparison, we tabulate SNRs of the noise-free cases computed with the aLIGO's design sensitivity PSD.}
    \centering
    \begin{tabularx}{1.\linewidth}{@{} l @{\extracolsep{\fill}} *{12}{c} @{}}
    \toprule
        \multirow{3}{*}{Noise} & \multicolumn{6}{c}{Homogeneous BBH pairs (template-target)} & \multicolumn{6}{c}{Heterogeneous BBH pairs (template-target)} \\
        \cline{2-7} \cline{8-13}
        & \multicolumn{3}{c}{Nonprecessing target}
        & \multicolumn{3}{c}{Precessing target}
        & \multicolumn{3}{c}{Nonprecessing target}
        & \multicolumn{3}{c}{Precessing target}\\
        \cline{2-4} \cline{5-7} \cline{8-10} \cline{11-13}
        & NU--NU & NU--NL & NL--NL & PU--PU & PU--PL & PL--PL & PU--NU & PU--NL & PL--NL & NU--PU & NU--PL & NL--PL \\
        \hline
        Free & $24.5$ & $27.2$ & $29.4$ & $35.6$ & $39.5$ & $42.7$ & $8.5$ & $10.9$ & $11.2$ & $12.3$ & $14.0$ & $16.3$ \\
        aLIGO & $24.1$ & $26.9$ & $29.2$ & $33.7$ & $37.5$ & $40.2$ & $11.1$ & $12.1$ & $5.4$ & $13.2$ & $15.3$ & $5.4$ \\
    \hline
    \hline
    \end{tabularx}
    \label{tab:snr}
\end{table*}

We can see from $s(t)$ and $|s(f)|$ in Fig.~\ref{fig:wf_noisy_aligo} that the presence of noise obviously hinders recognizing the difference shown in the noise-free $h(t)$ and $|h(f)|$. The difficulty in identifying the difference is also shown in the spectrogram presented in the bottom panel of Fig.~\ref{fig:wf_noisy_aligo}: We see the brightness---representing the energy content of the signal~\cite{Chatterji:2004qg}---of NU signal changes like PU signal as the time and frequency evolve. We also observe a few faint vertical nodes appear on both signals. 

From the spectrograms of NL and PL signals, we can observe changes in the brightness for $t\!<\!-0.1~\textrm{s}$ similar to NU and PU signals. However, we can observe two peaks---the primary peak at $t=0~\textrm{s}$ and the secondary peak at $t\sim-0.06~\textrm{s}$---agreeing with $\Delta t\simeq58~\textrm{ms}$ but not shown in the spectrograms of NU and PU signals. Despite this, the only recognizable differences between spectrograms of NL and PL signals are the amount of brightness affecting the apparent visibility of two peaks and the presence of a dark vertical node at different times, i.e., $t\!\sim\!-0.2~\textrm{s}$ for NL and $\sim\!-0.15~\textrm{s}$ for PL. 
Such small differences make it hard to identify whether a millilensed signal is a precessing or nonprecessing signal.

\subsection{Testing different template hypotheses via SNR}
\label{sec:SNR_comparison}

We compute the matched-filter SNR to quantitatively examine the difference. For this study, we configure different template-target pairs with the four types of signals studied thus far by taking whitened noise-free signals as templates and whitened noisy signals as targets. We classify the pairs into two categories, homogeneous BBH pairs and heterogeneous BBH pairs, defining homogeneous or heterogenous to represent whether the same nonprecessing or precessing binary is assumed for both template and target or not: Following the definition, `NX-NX' and `PX-PX' correspond to homogeneous pairs, regardless of `X' (either `U' or `L'), and `NX-PX' and `PX-NX' to heterogeneous pairs. We collect the pairs into two additional subcategories of nonprecessing or precessing targets. Given template and target signals, we use the \texttt{matched\_filter} function of the \texttt{pycbc.filter} package\footnote{https://pycbc.org/pycbc/latest/html/pycbc.filter.html} to compute the SNR and summarize the resulting SNRs in Table~\ref{tab:snr}.

From the SNRs of homogeneous pairs, we see SNRs for precessing targets are commonly $\sim\!40\%$ higher than corresponding SNRs for nonprecessing targets. We observe testing millilensed templates for millilensed targets shows the highest SNR among the pairs of each subcategory. We also find unlensed templates are applicable even for millilensed targets because $\textrm{SNR}_\textrm{NU-NL}\!>\!\textrm{SNR}_\textrm{NU-NU}$ and $\textrm{SNR}_\textrm{PU-PL}\!>\!\textrm{SNR}_\textrm{PU-PU}$. 

Obtaining higher SNR for precessing target is still shown from the SNRs of heterogeneous pairs. However, testing heterogeneous templates to given targets gives much lower SNRs than homogeneous pairs. In particular, as shown from PL-NL and NL-PL pairs in aLIGO noise, computing SNR with heterogeneous millilensed template is obviously disadvantageous for finding millilensed targets because $\textrm{SNR}_\textrm{PL-NL}$ and $\textrm{SNR}_\textrm{NL-PL}$ are lower than a conventional detection criterion $\geq\!8$ for the matched-filter SNR of a single detector~\cite{KAGRA:2013rdx}.

The result provides a conditional answer to our question: It is likely to distinguish NL signal from PU signal if an adequate template is tested and yields a higher SNR than other templates. For example, for NL target, we obtain SNRs as $29.2$, $26.9$, and $12.1$ with NL, NU, and PU templates, respectively. Similarly, we get SNRs for `PU' target as $33.7$ and $13.2$ with PU and NU templates, respectively. Moreover, discriminating a millilensed nonprecessing target from a millilensed precessing target is also possible if we compare SNRs of NL and PL targets, e.g., NL-NL ($29.2$) and PL-NL ($5.4$) or PL-PL ($40.2$) and NL-PL ($5.4$). However, it is important to note that this statement assumes that (i) the appropriate unlensed or lensed hypothesis is considered, and (ii) the correct intrinsic and extrinsic BBH parameters are used with and without the lensing effect, for the given target signal.


\subsection{Discriminating two effects via parameter estimation}
\label{sec:PE_example}

Following the matched-filter SNR computation, we conducted a PE study on the signals analyzed in the previous section. 
Four sets of signals -- $h_\textrm{NL}$, $h_\textrm{NU}$, $h_\textrm{PL}$, and $h_\textrm{PU}$ -- with parameters listed in Table~\ref{tab:params}, were injected into Gaussian noise of the aLIGO's design PSD~\cite{aligodesignpsd}.
\footnote{For comparison, the PSD data used for generating the strain data of the previous section is also known as \emph{zero-noise} PSD (e.g.,~\cite{Rodriguez:2013oaa, Favata:2021vhw}) that supposes no fluctuations around the given theoretical PSD.}
We used the \textsc{IMRPhenomXHM}~\cite{Garcia-Quiros:2020qpx} waveform approximant for the nonprecessing zero-spin GWs and \textsc{IMRPhenomXPHM}~\cite{Pratten:2020ceb} for generating the precessing GWs.
To model the lensed waveforms, we applied the phenomenological millilensing GW model developed in~\cite{Liu2023exploring}, which parametrizes the observables of millilensed GWs without assuming a specific lens mass distribution.
This approach allows us to map the results to the point-mass lens model utilized in the previous section.
Following~\cite{Liu2023exploring}, we express the lensed GW as 

\begin{equation}
    \tilde{h}_\textrm{ML}(f)\!=\! \tilde{h}_\textrm{U}(f) \sum_{j=1}^K \frac{d_L}{d^\mathrm{eff}_j} \exp(2 \pi i f  \Delta t_{j}\!-\!i\pi n_{j}),
    \label{eq:millilensed_waveform}
\end{equation}
where we sum over individual component lensed signals with $K$, a free parameter, representing the total number of component signals.
Each of the signals is described by the effective luminosity distance $d^\textrm{eff}_j \equiv d_L/\sqrt{\mu_{\textrm{rel},j}}$ for the given true luminosity distance to the source $d_L$ and relative magnification $\mu_{\textrm{rel},j}\!\equiv\!|\mu_j/\mu_1|$. 
The relative time delay of the $j^\textrm{th}$ signal with respect to the first lensed signal is expressed as  $\Delta t_j$, and $n_j$ denotes the Morse phase index of $j^\textrm{th}$ signal. 
For a point-mass lens, we inject $K\!=\!2$ signals with $n_1\!=\!0$ and $n_2\!=\!0.5$ and use a discrete uniform prior for $K\in\{1,2,3\}$ and $n_j\in\{0, 0.5, 1\}$, and uniform priors for $d^\textrm{eff}_j$ and $\Delta t_j$.

We performed a Bayesian inference of the BBH source and lensing parameters using the Bayesian inference library, \textsc{bilby}~\cite{bilby1, bilby2}. 
We used the nested sampling algorithm with the \textsc{pymultinest} sampler~\cite{skilling2006nested, Buchner2016pymultinest} 
setting the number of live points to $\textrm{nlive}=2048$.
The analysis was carried out with a minimum frequency of $f_\mathrm{min} = 20\,\mathrm{Hz}$ and a signal duration of $t=8\, \textrm{s}$.
The recovery was performed with \textsc{IMRPhenomXPHM} waveform approximant for all signals.

\begin{figure}[t]
    \centering
    \includegraphics[width=1\linewidth]{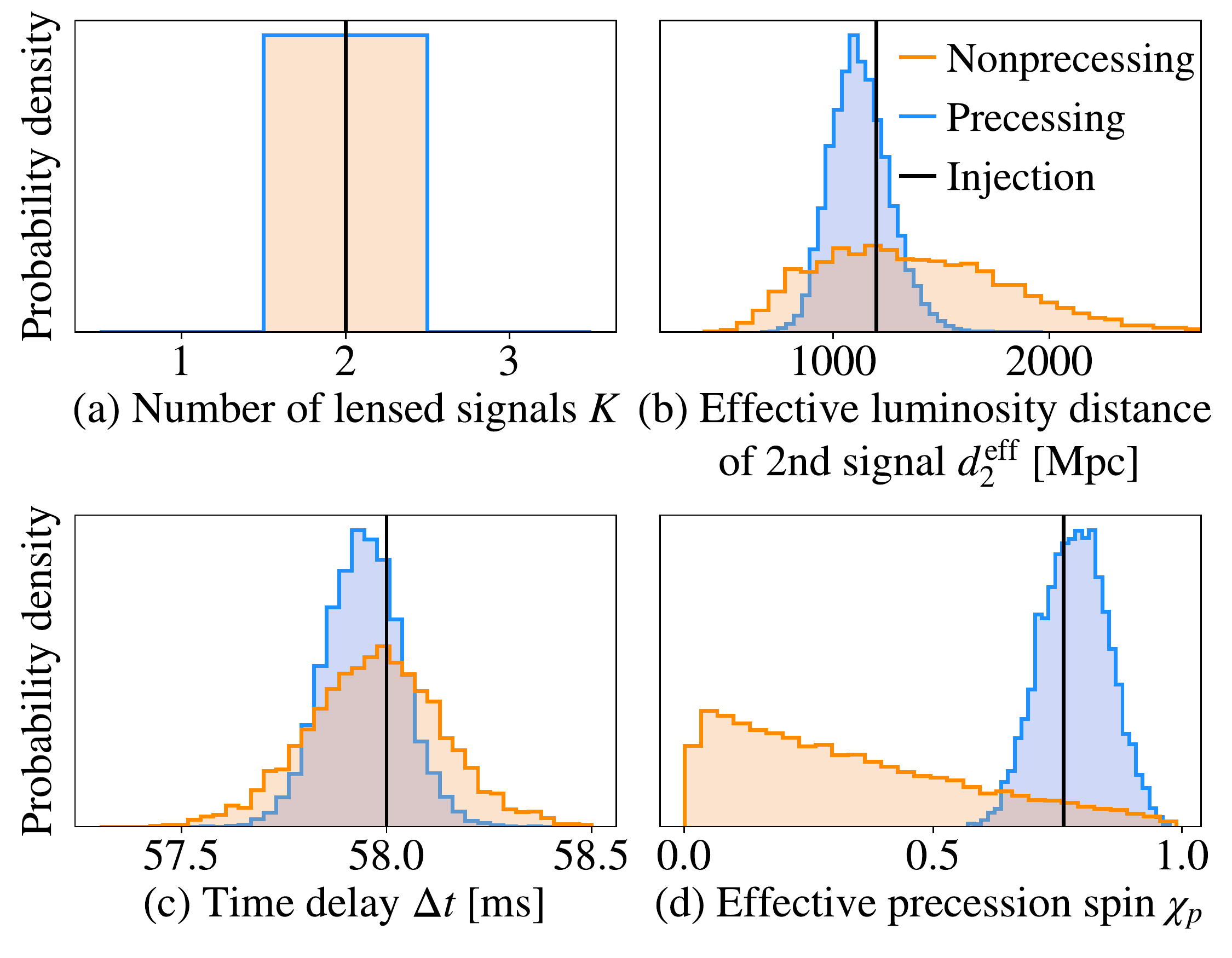}
    \caption{Posterior probability density functions (PDFs) of the parameters (a) number of component lensed signals $K$, (b) effective luminosity distance of the second signal $d_2^\mathrm{eff}$, (c) relative time delay between the two signals $\Delta t$, and (d) effective precession spin $\chi_p$ with the vertical line indicating the injected spin of the precessing GW (for the nonprecessing injection, $\chi_p$ is set to zero).
    The precessing and nonprecessing signals lead to consistent results with the injected values (vertical black lines). 
    We conclude that in this example we can correctly recover both lensing and precession in the presence of both effects.}
    \label{fig:pdfs}
\end{figure}

\begin{figure}
    \centering
    \includegraphics[width=\linewidth]{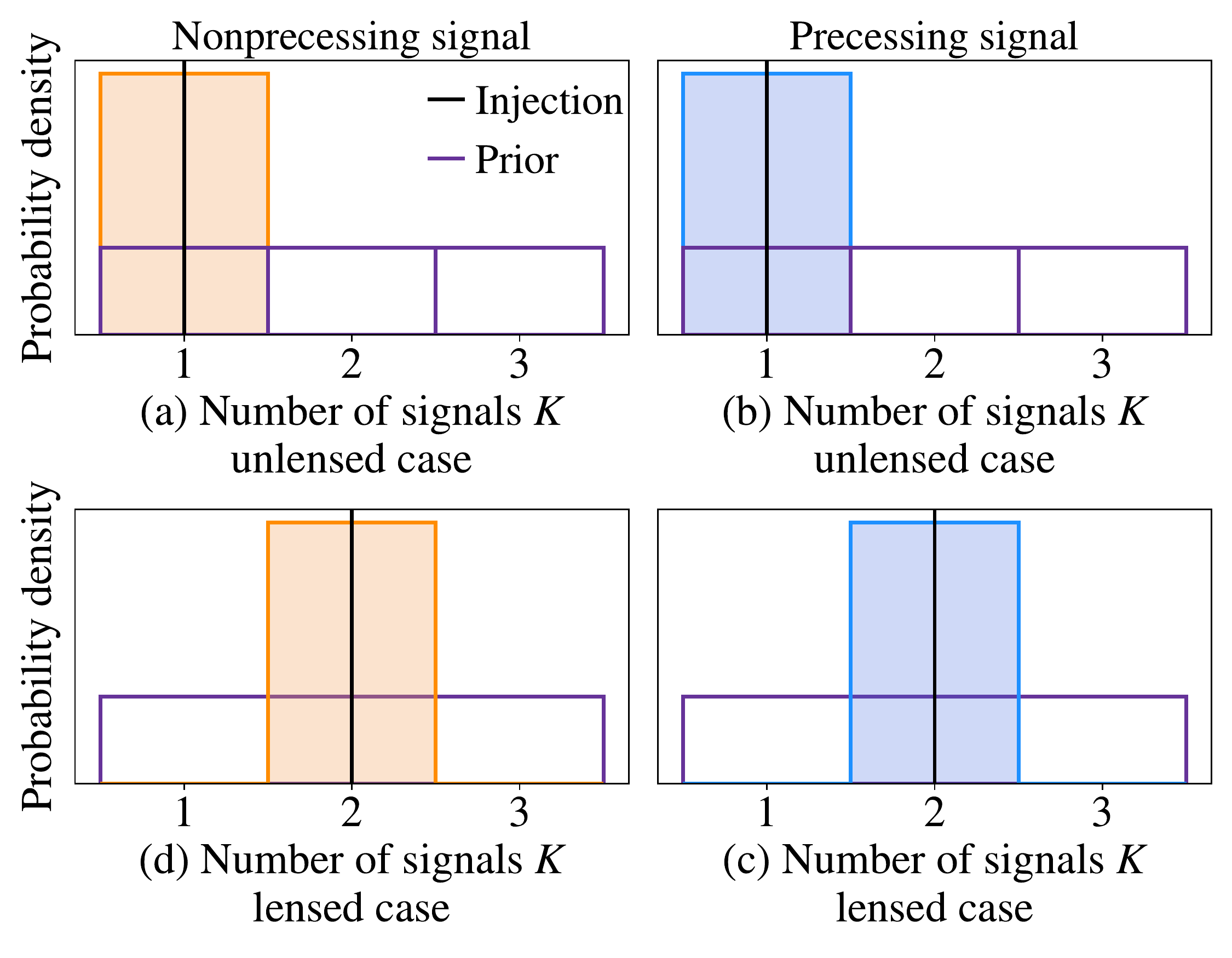}
    \caption{Recovery of the number of component signals $K$ for different cases: (a) nonprecessing unlensed (NU), (b) precessing unlensed (PU), (c) nonprecessing lensed (NL), and (d) precessing lensed (PL) signals.
    The left column shows the expected results for nonprecessing cases, where the number of component signals is accurately recovered, as there are no effects that could be degenerate with lensing.
    The right column presents the precessing cases, confirming that precession does not interfere with the correct identification of $K$.
    This holds true for both lensed signals with $K=2$ and unlensed signals with $K=1$.
    In all cases, $K$ was sampled over a discrete uniform prior $K\in(1,2,3)$ allowing for up to three overlapping signals.} \label{fig:illustr_example_number_of_signals_K}
\end{figure}

In Fig.~\ref{fig:pdfs}, we present the PE results for both the PL and NL signals. 
While a full PE was performed over all parameters, for the clarity of the figure, we show only the results of the selected lensing and precession parameters.
We correctly recover the number of lensed signal components $K=2$, for both the precessing and nonprecessing lensed GWs (Fig.~\ref{fig:pdfs} (a)). 
The posteriors for $d^\textrm{eff}_2$ and $\Delta t$ correctly recover the injected values, marked by vertical black lines, with $d^\textrm{eff}_2\!\simeq\!1200~\textrm{Mpc}$ and $\Delta t\!\simeq\!58~\textrm{ms}$, for both precessing and nonprecessing cases (Figs.~\ref{fig:pdfs} (b) and (c), respectively).
Additionally, the recovered $\chi_p$ posterior distribution agrees with the injected value of $\chi_p= 0.76$ for the lensed precessing signal, and rails toward $\chi_p= 0$ for the nonprecessing signal, in agreement with the injected zero spin (Fig.~\ref{fig:pdfs} (d)).
This example confirms that, for the analyzed signals, the identification of the injected millilensed GWs via PE recovery is not impacted by the presence of precession. 
Specifically, we can correctly identify the number of lensed signals $K$, which serves as a criterion for identifying millilensing throughout this study.
Furthermore, the results demonstrate that, for the chosen set of parameters, we can successfully identify a precessing GW signal even in the presence of millilensing.

Moreover, in Fig.~\ref{fig:illustr_example_number_of_signals_K}, we present the recovery of the number of signals $K$ for four cases: NL, NU, PL, and PU.
In these scenarios, the injected lensed signals had $K=2$, while the injected unlensed signals had $K=1$ component. 
All sampling was performed over a discrete uniform prior $K\in\{1,2,3\}$.
The results indicate that, regardless of the presence or absence of precession, the number of component signals is accurately recovered. 
This suggests that, for the given set of parameters, the lensing effects can be correctly identified using the number of component signals as the identification criterion.

We can observe from Figs.~\ref{fig:pdfs} (b) and (c) that the width of the posterior for precessing signal is much narrower than the nonprecessing one.
This is due to the different SNRs of two signals, $\textrm{SNR}^\textrm{PE}_\textrm{PL-PL}\!=\!30.5\!>\!\textrm{SNR}^\textrm{PE}_\textrm{NL-NL}\!=\!15.6$.
For a detailed study, we simulate another set of signals to probe the possible biases arising from the two effects, which we present in the following section.
\section{Which signals produce the most bias? Probing a wider parameter space}
\label{sec:PE_broad_study}

The example presented in the previous section suggests that we can correctly identify the effects of both gravitational millilensing and precession when they occur simultaneously  for a 30$M_\odot$-10$M_\odot$ BBH system. 
We further explore a broader parameter space to draw more definitive conclusions on the distinguishability of the two effects and study potential degeneracies for different signals.
We perform an injection study by simulating lensed precessing GWs and injecting them into ground-based LIGO-Virgo detectors with O4 sensitivity ~\cite{aasi2015advanced, O4ligodesignpsd} and zero noise.
We study different values of source masses, spins, inclination angles, and SNRs, as listed in Table~\ref{tab:injection_params_PE}.
We then recover these injected signals with three different models: (i) precessing lensed (PL), (ii) precessing unlensed (PU), and (iii) nonprecessing lensed (NL). 
In the NL recovery, we adopt a nonprecessing waveform approximant \textsc{IMRPhenomXHM} and assume aligned spin priors for the BBH spin components.
The PE study can help us determine the signals for which the confusion between gravitational lensing of GWs and spin-induced precession is most significant.

\renewcommand{\cellalign}{cl}

\begin{table}[]
    \centering
    \caption{Summary of BBH injection parameters varied across the simulated injected signals. We vary selected parameters at a time for each subsection, as explained for specific cases in the main text. We control varying lensing strength by altering the relative source position parameter $y$ and control varying precession by altering the component spins and inclination angles (representing the angle between the total angular momentum $\Vec{J}$ and the line of sight $\Vec{N}$), which lead to different values of the effective precessing spin $\chi_p$.}
    \begin{tabularx}{1.\linewidth}{l @{\extracolsep{\fill}} c @{\extracolsep{\fill}} c}
        \toprule
         Parameters & Unit & Values \\
         \hline
         Component masses, ($m_1, m_2$) & $M_\odot$ & (30, 10), (14, 10), (30, 35) \\
         \multirow{2}{*}{Spin magnitudes, ($a_1, a_2$)} & \multirow{2}{*}{--} & (0.43, 0.35), (0.58, 0.56), \\
         & & (0.78, 0.82) \\
         Precessing spin, $\chi_p$ & -- & 0.16, 0.36, 0.64 \\
         Inclination angle, $\theta_{JN}\sphericalangle{\Vec{J} \Vec{N}}$ & rad & 0.5, 0.55, 0.6 \\
         Relative source position, $y$ & -- & 0.1, 0.3, 0.5 \\
         SNR & -- & 8, 18, 40 \\
        \hline
        \hline
    \end{tabularx}
    \label{tab:injection_params_PE}
\end{table}

\begin{table}[]
    \centering
    \caption{Log Bayes factors, $\ln \mathcal{B}$, relative to the noise of the recovered signals with three different models: PL - lensed and precessing, NL - nonprecessing lensed, PU - precessing unlensed.
    The injection is lensed and precessing GW signal, with varying precession spin among the 3 values listed in the first column. 
    The injected masses are $(14, 10)M_\odot$ and SNR was set to 18 in the injection.
    For all three values of spin, the Bayes Factor indicates nonprecessing lensed (NL) recovery to be the most favored model, although it does not differ significantly from the PL recovery. 
    Notably, the precession-only recovery is disfavoured for all three spin values.
    Interestingly, one might expect the BF for PL recovery to be smaller or less favorable compared to the PU recovery due to the larger number of parameters involved (Occam's razor).
    However, it still shows a preference for PL over the PU recovery.
    This suggests that the GW signals are predominantly influenced by the lensing effects rather than precession.
    These results correspond to the results of injection runs summarized in Table~\ref{tab:summary_vary_spin}.}
    \begin{tabularx}{1.\linewidth}{l  @{\extracolsep{\fill}} c @{\extracolsep{\fill}}  @{\extracolsep{\fill}} c}
        \toprule
         Injected spin $\chi_p$ & Recovery model & $\ln \mathcal{B}$ \\
         \hline 
         \multirow{3}{*}{0.16} & PL & $112.742 \pm 0.213$  \\
         & NL & $114.550 \pm 0.205$   \\
         & PU & $72.745 \pm 0.205$   \\
         \hline 
         \multirow{3}{*}{0.36} & PL & $112.665 \pm 0.210$  \\
         & NL & $113.326 \pm 0.204$  \\
         & PU & $70.756 \pm 0.173$  \\
         \hline 
         \multirow{3}{*}{0.64} & PL & $108.567 \pm 0.214$  \\
         & NL & $109.903 \pm 0.205$  \\
         & PU & $68.270 \pm 0.205$  \\
        \hline
        \hline
    \end{tabularx}
    \label{tab:BFs}
\end{table}

We compare the Bayes' factors (BFs) recovered from the PE study in Table~\ref{tab:BFs}. 
From the comparison of NL and PU models, we can see that lensing is preferred, with the relative BF between two models $\ln \mathcal{B}^{\textrm{len}}_{\textrm{prec}} \equiv \ln \mathcal{B}_{\textrm{len}} - \ln \mathcal{B}_{\textrm{prec}} > 41$ for all three spin values.
This suggests that lensing-only recovery is significantly preferred over precession-only recovery even for highly-spinning BBH.
We present the recovery of lensing and precession parameters in the following sections.

\begin{figure}
    \centering
    \includegraphics[width=\columnwidth]{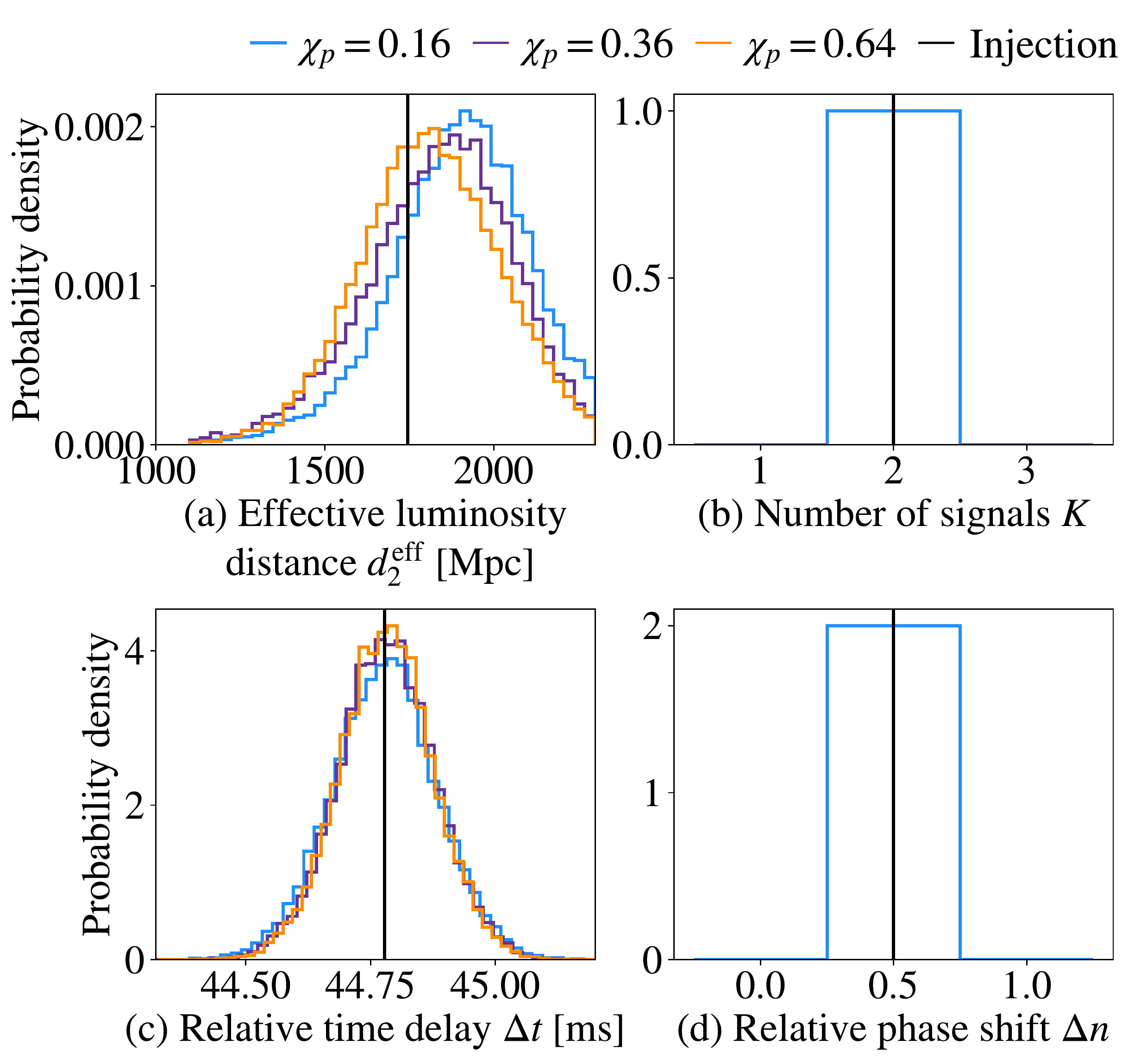}
    \caption{Posterior distribution of (a) the effective luminosity distance $d_2^\mathrm{eff}$ of the second component GW signal, (b) number of component lensed signals $K$, (c) relative time delay $\Delta t$ and (d) relative phase shift $\Delta n$ between two component lensed GWs, recovered from injections with lensed and precessing waveform model (\textsc{IMRPhenomXPHM}) with small masses $(14, 
    10) M_\odot$, SNR 18 and varying effective precessing spin $\chi_p$. The vertical black lines represent the injected values.
    As the effective precessing spin decreases, there is an observed shift of the $d_2^\mathrm{eff}$ posterior distribution toward higher values. 
    The recovered distribution of the relative time delay $\Delta t$, number of component signals $K$ and relative phase shift $\Delta n$ do not show any significant dependence on the effective precessing spin. The posterior distribution of $K$ and $\Delta n$ are identical for three spin cases.
    These results provide important validation for lensing recovery, demonstrating outcomes supporting the presence of lensing despite the presence of precession in the recovery model.}
   \label{fig:dL_dt_small_mass_vary_spin}
\end{figure}

\begin{figure}
    \centering
    \includegraphics[width=\columnwidth]{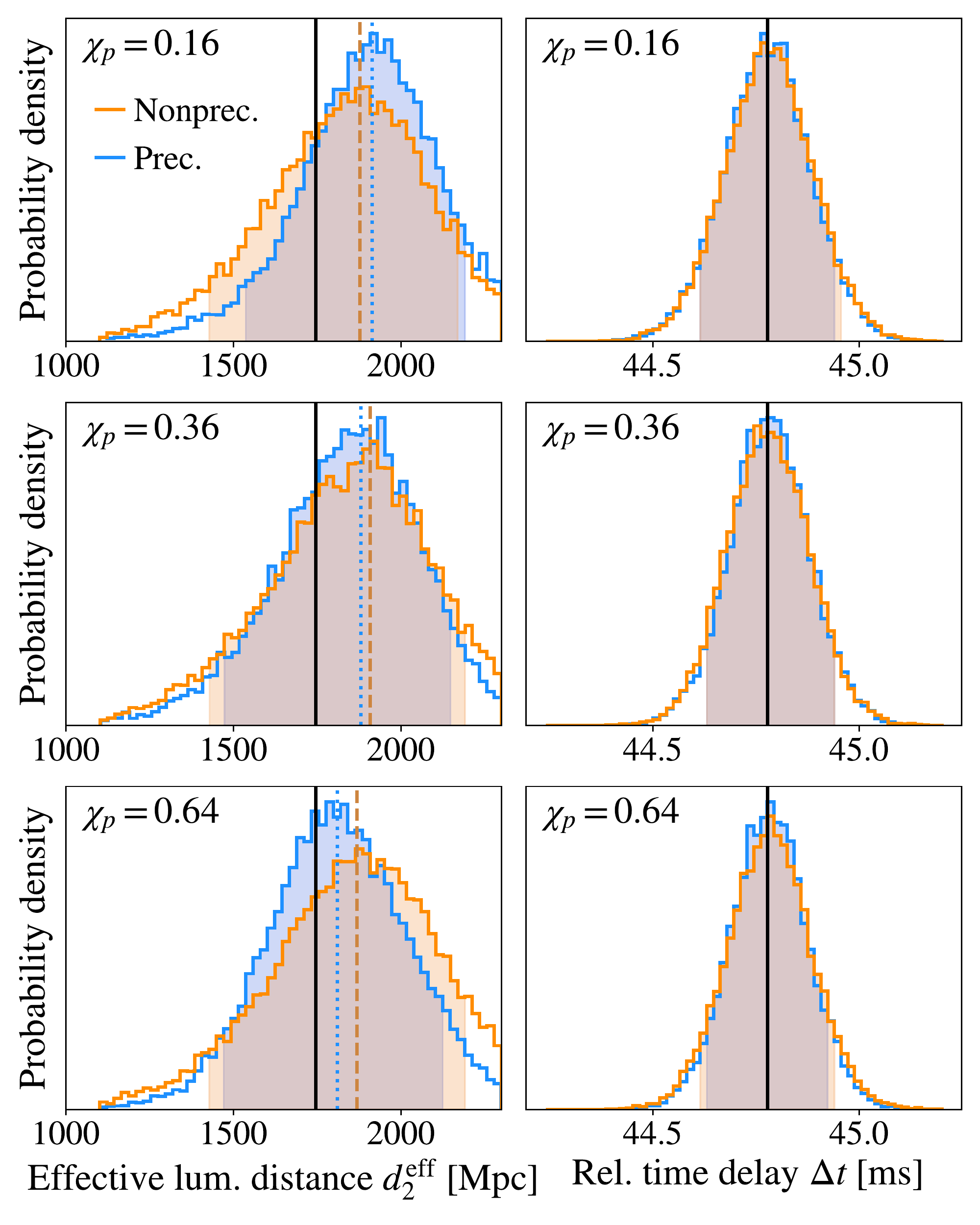}
    \caption{Posterior distribution of the effective luminosity distance $d_2^\mathrm{eff}$ (\textit{left column}) of the second component GW signal and relative time delay $\Delta t$ (\textit{right column}), recovered from injection with slowly-precessing $\chi_p = 0.16$ (\textit{top}), moderately-precessing $\chi_p = 0.36$ (\textit{middle}) and highly-precessing systems $\chi_p = 0.64$ (\textit{bottom}), with small masses $(14, 
    10) M_\odot$ and an SNR of 18. 
    The results present recovery with two models: lensed precessing (blue) and lensed nonprecessing (orange).
    The offset between the peaks (dashed and dotted lines) obtained from estimating kernel density of the recovered precessing and nonprecessing $d_2^\mathrm{eff}$ distributions is largest for highly precessing case, equivalent to 58 Mpc for the $\chi_p = 0.64$ recovery, 27 Mpc for $\chi_p = 0.36$ and 35 Mpc for $\chi_p = 0.16$. 
    In the highest-spinning $\chi_p = 0.64$ case, the precessing recovery is consistent with injected value, whereas the nonprecessing recovery peaks at higher values, although the injected value remains within 90\% C.I. represented by shaded regions. 
    Notably, the recovery of $\Delta t$ does not show any dependence on precession.
    }
    \label{fig:deff2_dt_prec_vs_nonprec}
\end{figure}

\begin{figure}[t!]
    \centering
    \includegraphics[width=0.9\columnwidth]{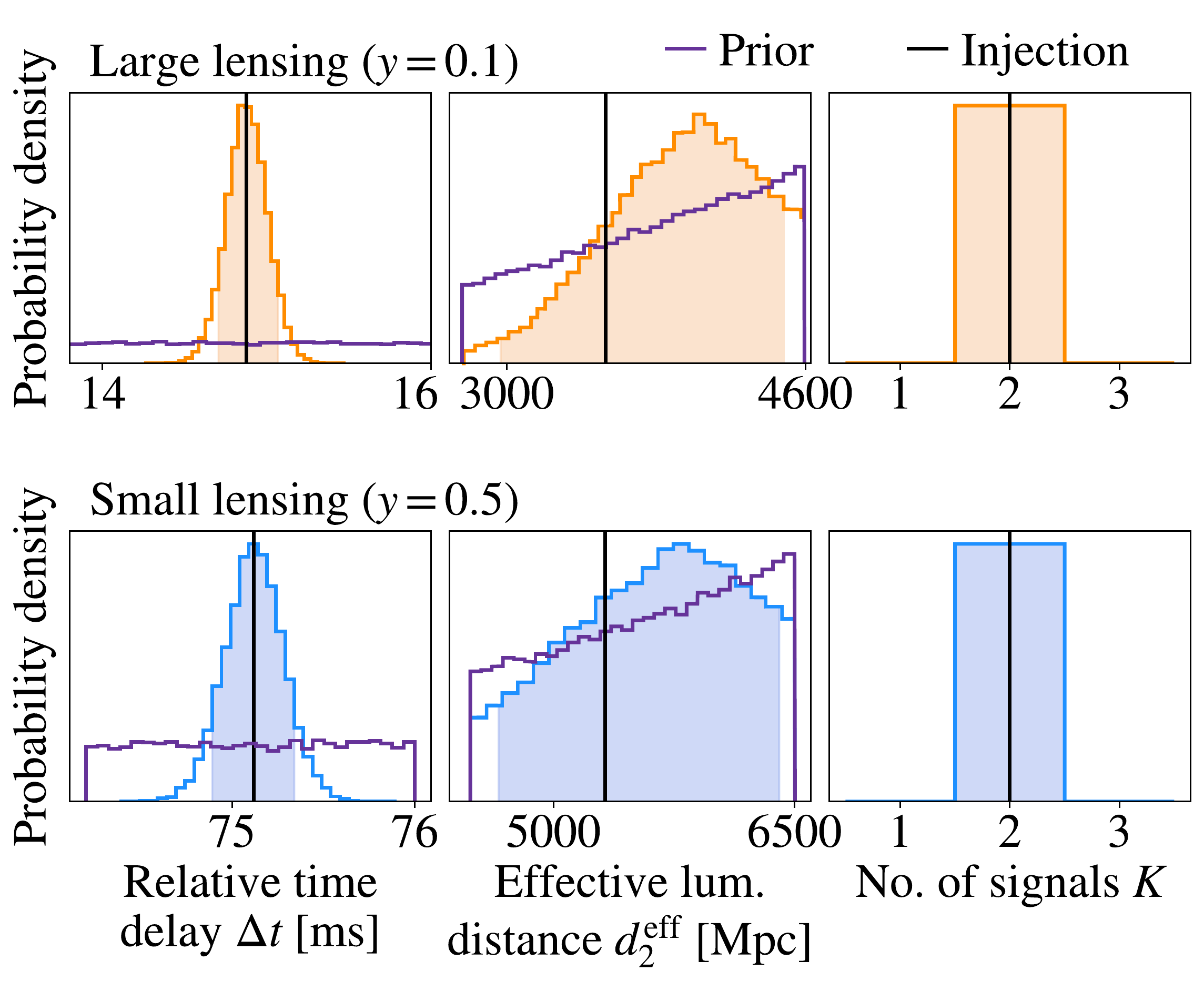}
    \caption{Recovery of lensing parameters for a high mass $(30, 35) M_\odot$, highly spinning $\chi_p = 0.64$ system, injected with an SNR of 18.
    The results are compared for two lensing scenarios: large $y=0.1$ (\textit{top}) and small $y=0.5$ (\textit{bottom}).
    In both cases, the time delay is recovered in agreement with the injection value, while the effective luminosity distance posterior shows a preference for higher values than injection.
    The prior distributions are shown for reference.
    The number of recovered component lensed signals $K$, shown in the last column, aligns with the injected scenario of $K=2$ millilensed signals, confirming the correct identification of lensing in the signal.}
    \label{fig:snr18_large_vs_small_lens}
\end{figure}

\subsection{Recovery of lensing parameters with and without precession}

In this section, we focus on the lensing parameters of the recovered lensed and precessing signals with varying effective precessing spins $\chi_p = (0.16, 0.36, 0.64)$. 
In this analysis, the lensed and precessing recovery is performed using the same model and waveform approximant (\textsc{IMRPhenomXPHM}) as the injection.
The goal is to assess whether the presence of precession affects the correct recovery of gravitational lensing parameters.
We find that the lensing parameters can be recovered consistently with the injected values, as shown in Fig.~\ref{fig:dL_dt_small_mass_vary_spin}. 
However, in the recovery of the effective luminosity distance (Fig.~\ref{fig:dL_dt_small_mass_vary_spin} (a)), there is an observable trend preferring larger effective luminosity distance values for smaller precessing spins. 
We have investigated changing the effective luminosity distance prior from power law to uniform, but the trend continued to remain (see Fig.~\ref{fig:dL2_prior_comparison} in Appendix~\ref{sec:appendix_priors}), suggesting it is not a direct consequence of the prior choice. 
In the recovery of the relative time delay $\Delta t$, the number of component signals $K$ and relative phase shift $\Delta n$, no dependence on the effective spin is observable.
All three simulations lead to consistent results with the injected value of $\Delta t$, $K$ and $\Delta n$, as shown in Fig.~\ref{fig:dL_dt_small_mass_vary_spin}.

We perform a recovery of the same lensed precessing signals, this time assuming they are only lensed but nonprecessing. 
We aim to test whether neglecting precession leads to observable biases in the information carried by lensing parameters. 
For that, we use the \textsc{IMRPhenomXHM} waveform approximant.
We find that despite neglecting the precession and using a wrongly assumed nonprecessing model in the recovery, we can still obtain results consistent with injected values within 90\% credible interval (C.I.) (see Fig.~\ref{fig:deff2_dt_prec_vs_nonprec}). 
As can be noted from Fig.~\ref{fig:deff2_dt_prec_vs_nonprec}, the nonprecessing recovery of $d_2^\mathrm{eff}$ (orange distribution) does not vary across the different spin injections. 
In contrast, the precessing recovery prefers higher values than the injected $d_2^\mathrm{eff}$ at the lowest spin, but peaks near the injected value for the high-spin case.
This suggests that neglecting precession does not lead to significant biases for low and moderate spins ($\chi_p\sim 0.4$), but can result in biases in the effective luminosity distance recovery for spins of the order of $\chi_p\sim 0.6$.
In contrast, the recovery of time delay $\Delta t$ does not show significant differences between precessing and nonprecessing recovery models.
We, therefore, conclude that the information about the relative time delay between lensed signals is not affected by the presence of precession, even for highly spinning systems.

The results presented so far assumed medium-lensing strength ($y=0.3$).
We also performed an injection study of highly precessing lensed GWs with varying lensing strength.
We control the strength of lensing by varying the value of the relative source position $y\in(0.1, 0.3, 0.5)$, corresponding to large, medium, and small lensing effects, respectively.
We then convert this into the observable lensing parameters $d_2^\mathrm{eff}, \Delta t, n$, which quantify the lensing magnification, time delay, and phase shift, respectively.
We present the posterior distributions for the effective luminosity distance and the relative time delay in Fig.~\ref{fig:snr18_large_vs_small_lens}.
For both small and large lensing cases, the relative time delay $\Delta t$ is recovered in agreement with the injected values. 
However, the effective luminosity distance posterior $d_2^\mathrm{eff}$ is not fully informative, with a preference for higher values and railing against the prior edges, once again suggesting that lensing magnification can be biased for highly spinning BBH systems.

\begin{figure}[t!]
    \centering
    \includegraphics[width=1\linewidth]{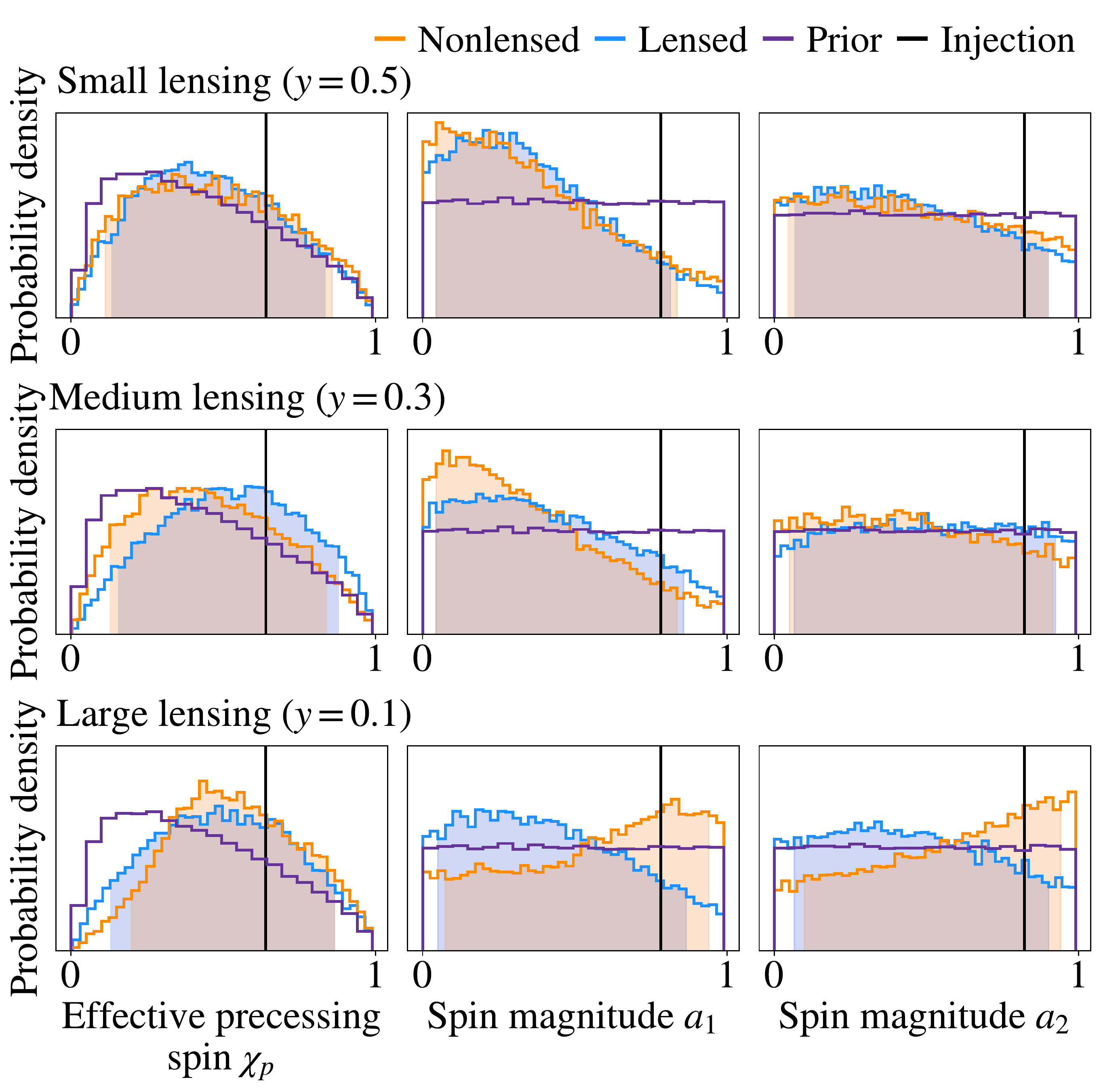}
    \caption{Recovery of the effective precessing spin $\chi_p$ and spin magnitudes $a_1$, $a_2$ for high mass $(30, 35) M_\odot$, high spin $\chi_p=0.64$, SNR of 18 lensed precessing injections (vertical black lines) with varying lensing strength from small (\textit{top}) to large lensing (\textit{bottom}). Recovery with and without lensing is shown in blue and orange, respectively. The shaded regions correspond to the 90\% credible intervals. We find that for small and medium-lensing, neglecting the lensing effect does not affect the recovered spins significantly. 
    However, it becomes important in the large lensing scenario.}
    \label{fig:component_spins_vary_lens}
\end{figure}

\subsection{Recovery of precession parameters with and without lensing}

We also study the ability to recover the effective precessing spin $\chi_p$ in the presence of lensing with varying degrees of lensing effect. 
We present the recovery of the effective precessing spin $\chi_p$ for a high-mass, high-spin injection in Fig.~\ref{fig:component_spins_vary_lens}, showing the comparison of recovery with and without lensing. 
Since the effective precessing spin $\chi_p$ is not a parameter directly inferred from the data, but computed using the component spins, we also show the recovered dimensionless spin distributions $a_1$ and $a_2$ in Fig.~\ref{fig:component_spins_vary_lens}.
For the effective precessing spin, it can be noted that the lensed and unlensed recovery do not lead to significant differences. 
In particular, for the small-lensed case, the two recovery distributions are almost identical (upper left panel in Fig.~\ref{fig:component_spins_vary_lens}).
For the medium- and large-lensed cases, the distribution shape shows small differences, with the injected value remaining within 90\% C.I. in both cases. 
The results for component spins demonstrate more clearly how the recovery between lensed and unlensed cases differ with increasing lensing strength. 
Although the posterior distributions of $a_1$ and $a_2$ are railing toward lower and upper edges in all cases, it is clear that in the large-lensing scenario, the recovery with lensed signal tends toward smaller spin values, while the unlensed recovery rails against extreme spins. 
This suggests that the increased spins compensate for the neglected lensing effect in the large lensing $y=0.1$ case.
Overall, we conclude that neglecting the lensing effect does not significantly affect the recovery of spins for small- to medium-lensing effects. 
However, for large lensing, including the lensing can make a significant difference in the recovered spin information. 

\begin{figure}[t!]
    \centering
    \includegraphics[width=\columnwidth]{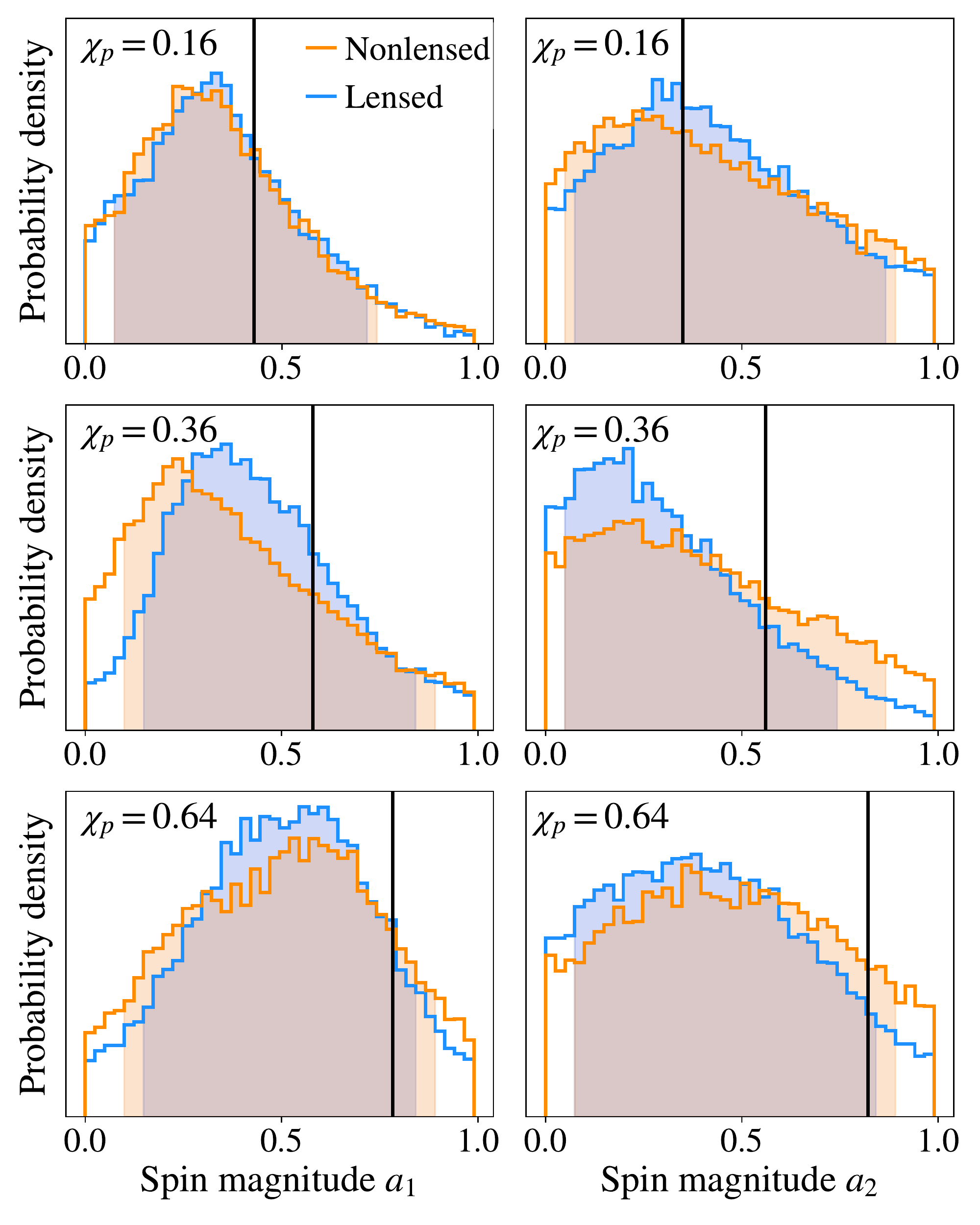}
    \caption{Comparison of lensed (blue) and nonlensed (orange) recovery of spin magnitudes for slowly $\chi_p = 0.16$ (\textit{top}), moderately $\chi_p = 0.36$ (\textit{middle}) and highly-spinning $\chi_p = 0.64$ (\textit{bottom}) systems.
    The injection was performed with small masses $(14, 
    10) M_\odot$, medium-lensing $y=0.3$ and SNR of 18.
    The shaded regions correspond to 90\% C.I. of the posteriors. The posteriors are not informative, making it difficult to conclude the effect of lensing. From the comparison of lensed and unlensed distributions, lensing does not appear to significantly affect the posterior distribution. 
    For both $a_1$ and $a_2$, the high-spinning case does not peak near the injected value. However, the impact of lensing on highly spinning scenarios cannot be confirmed from this result. 
    The absence of lensing does not lead to significant differences in the posteriors.}
    \label{fig:a1_a2_slow_high_spin}
\end{figure}

We also compare the recovery of dimensionless spins for low- to high-spin medium-lensing injections, recovered with and without lensing, for smaller component masses (14, 10 $M_\odot$) where we expect the precessing and lensing effect to have accumulated the most. 
The reason we choose the medium-lensing ($y=0.3$) case, is that, while the large-lensing case ($y=0.1$) was identified as particularly interesting in Fig.~\ref{fig:component_spins_vary_lens}, we acknowledge that it represents an extreme scenario, situated at the boundary between the geometric and wave optics approximations~\cite{Liu2023exploring}. 
Hence, to provide a more balanced analysis, we opted to use a medium-lensing in the following.
As can be seen in Fig.~\ref{fig:a1_a2_slow_high_spin}, the posterior distributions do not show significant differences between the lensed and unlensed recoveries. 
Therefore, we conclude that the absence of lensing does not have a significant impact on the recovery of dimensionless spins in a medium-lensing scenario, even for a longer-duration signal, regardless of the injected spin values. 

\subsection{Comparing different SNRs}

\begin{figure}[t!]
    \centering
    \includegraphics[width=1\linewidth]{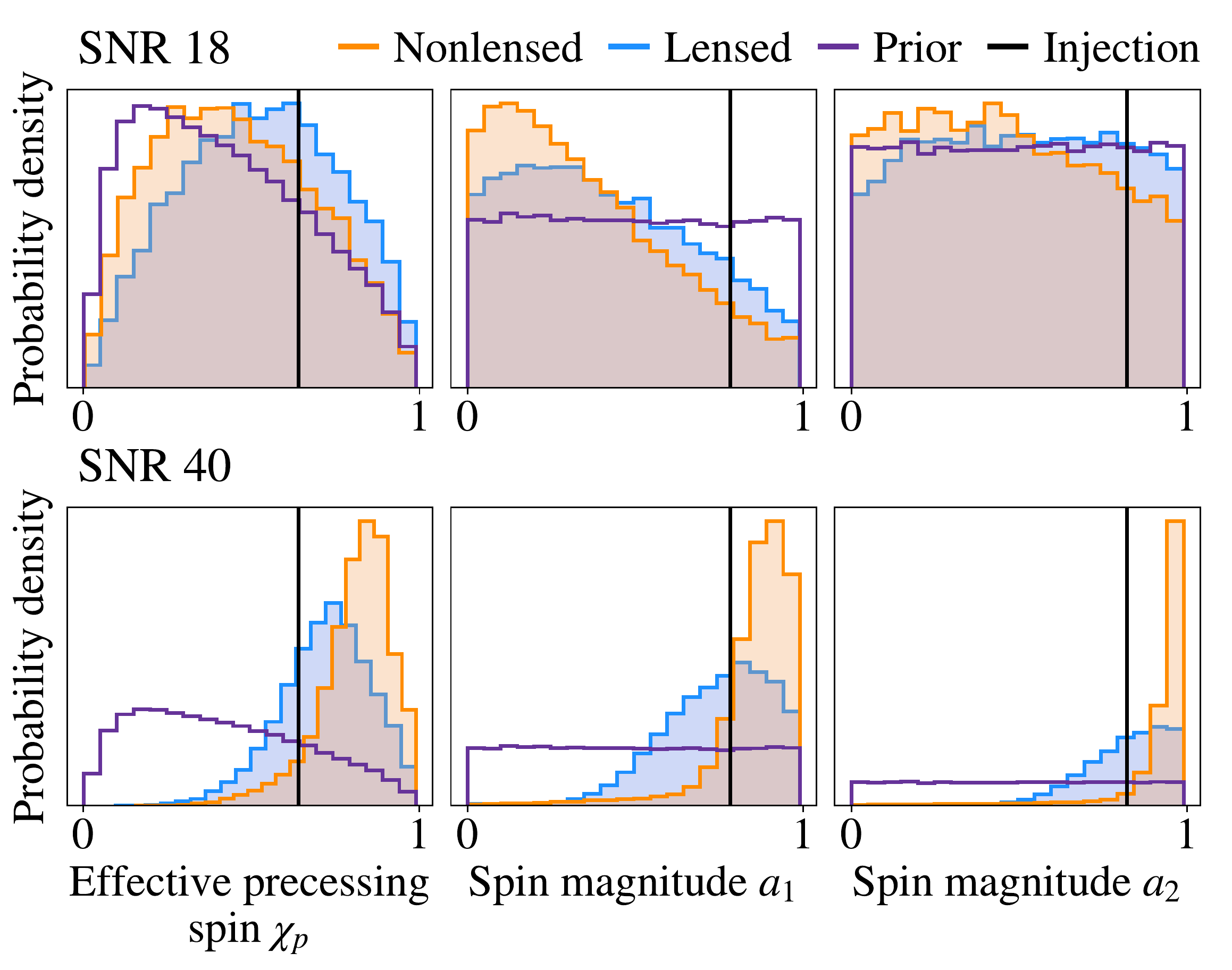}
    \caption{Comparison of SNR 18 (\textit{top}) and SNR 40 (\textit{bottom}) recovery of spin parameters with and without accounting for lensing (blue and orange, respectively) from a high mass $(30, 35) M_\odot$, high spin $\chi_p=0.64$ and medium-lensing $y=0.3$ injection. 
    The results for SNR 40 clearly demonstrate that including lensing in the recovery model (blue distribution) aligns well with the injected values. 
    In contrast, the recovery without lensing shows a pronounced preference for extreme spins, indicating a bias in parameter estimation.
    For SNR 18, the spin magnitude posteriors are notably less informative, displaying a tendency toward smaller spins than those injected.
    This highlights the importance of lensing in accurately recovering spin parameters, particularly in lower SNR cases, where the absence of lensing leads to misleading conclusions.}
    \label{fig:snr18_snr14_spins}
\end{figure}

\begin{figure}[t!]
    \centering
    \includegraphics[width=1\linewidth]{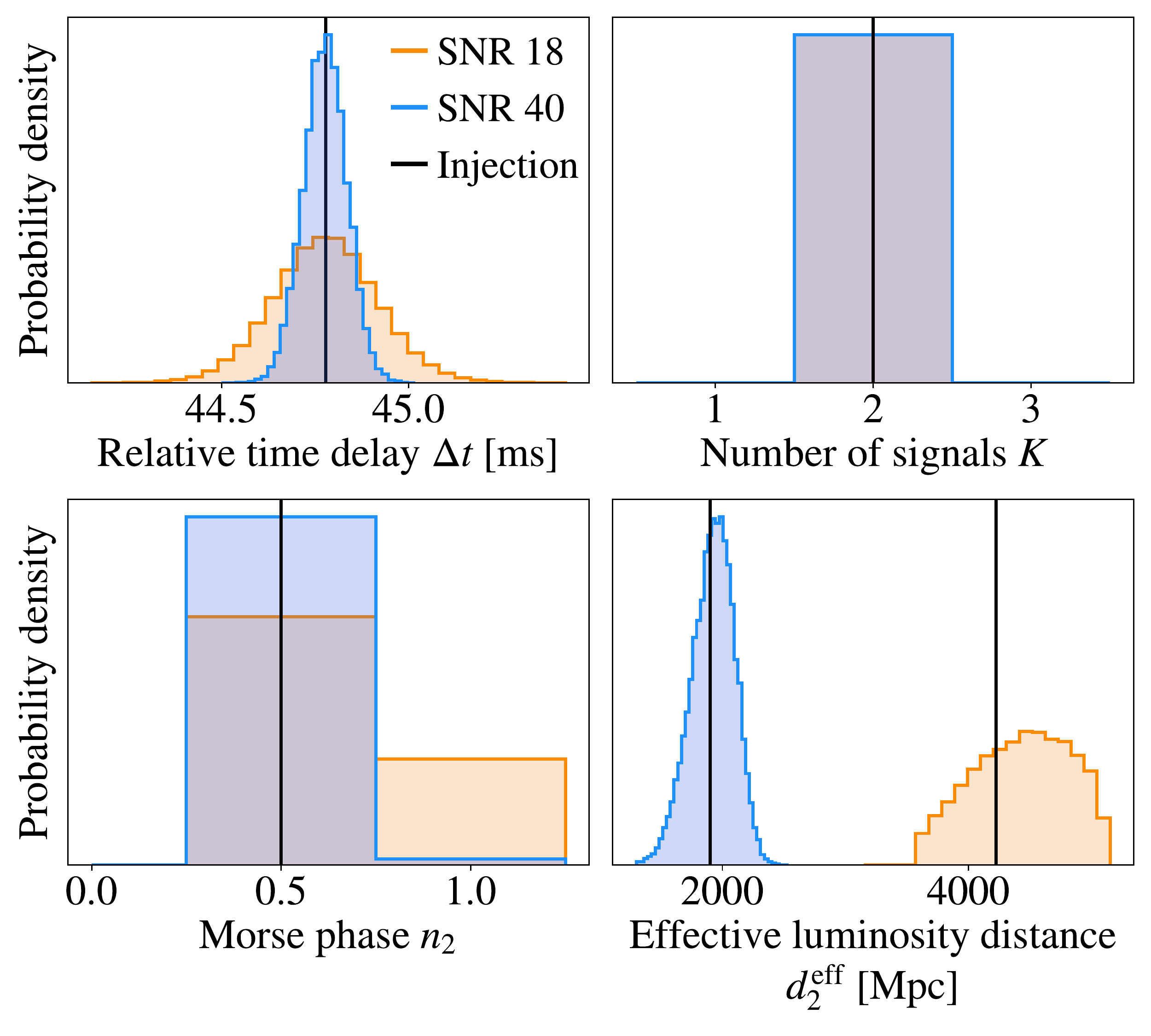}
    \caption{Comparison of the recovery of lensing parameters in the presence of precession at SNR 18 (orange) and SNR 40 (blue), corresponding to injection with high masses $(30, 35) M_\odot$, high spin $\chi_p=0.64$ and medium-lensing $y=0.3$.
    For both SNRs, the recovered posteriors agree well with the injection values represented by vertical black lines.
    The SNR of the injection was scaled by scaling the luminosity distance and the effective luminosity distance, hence the two runs have different values of the injected effective luminosity distance $d^{\rm{eff}}_2$ shown in the last plot.}
    \label{fig:snr18_snr40_comparison}
\end{figure}

Finally, we performed injections with different SNRs. 
We scaled the injected SNR of each signal by adjusting its injected luminosity distance accordingly.
In Fig.~\ref{fig:snr18_snr14_spins}, we present a comparison of the spin recovery for large-mass, highly spinning injections with SNR 18 and 40.
By comparing the dimensionless spin components recovery, it can be noted that the higher SNR case (SNR 40) leads to results consistent with injected values when lensing is included in the recovery. 
Neglecting lensing, however, demonstrates a tendency towards extremal spins for the SNR 40 case. 
For SNR 18 injection, the lensed recovery of the effective precessing spin agrees with the injected value, while the dimensionless spins are not informative. 
We also present the results of lensing parameters obtained with SNR 18 and SNR 40 (see Fig.~\ref{fig:snr18_snr40_comparison}). 
The results for the SNR 18 and SNR 40 cases are consistent with each other, both recovering injected lensing parameters correctly. 
We conclude that, for signals with SNRs of 18 and higher, we can correctly recover the lensing parameters in the presence of precession.
However, an SNR of 18 is not sufficient to accurately infer dimensionless spin components for a lensed and precessing signal.
Even at a high SNR of 40, neglecting lensing in the recovery model can bias the inferred spin information.

\begin{figure}[t!]
    \centering
    \includegraphics[width=1\linewidth]{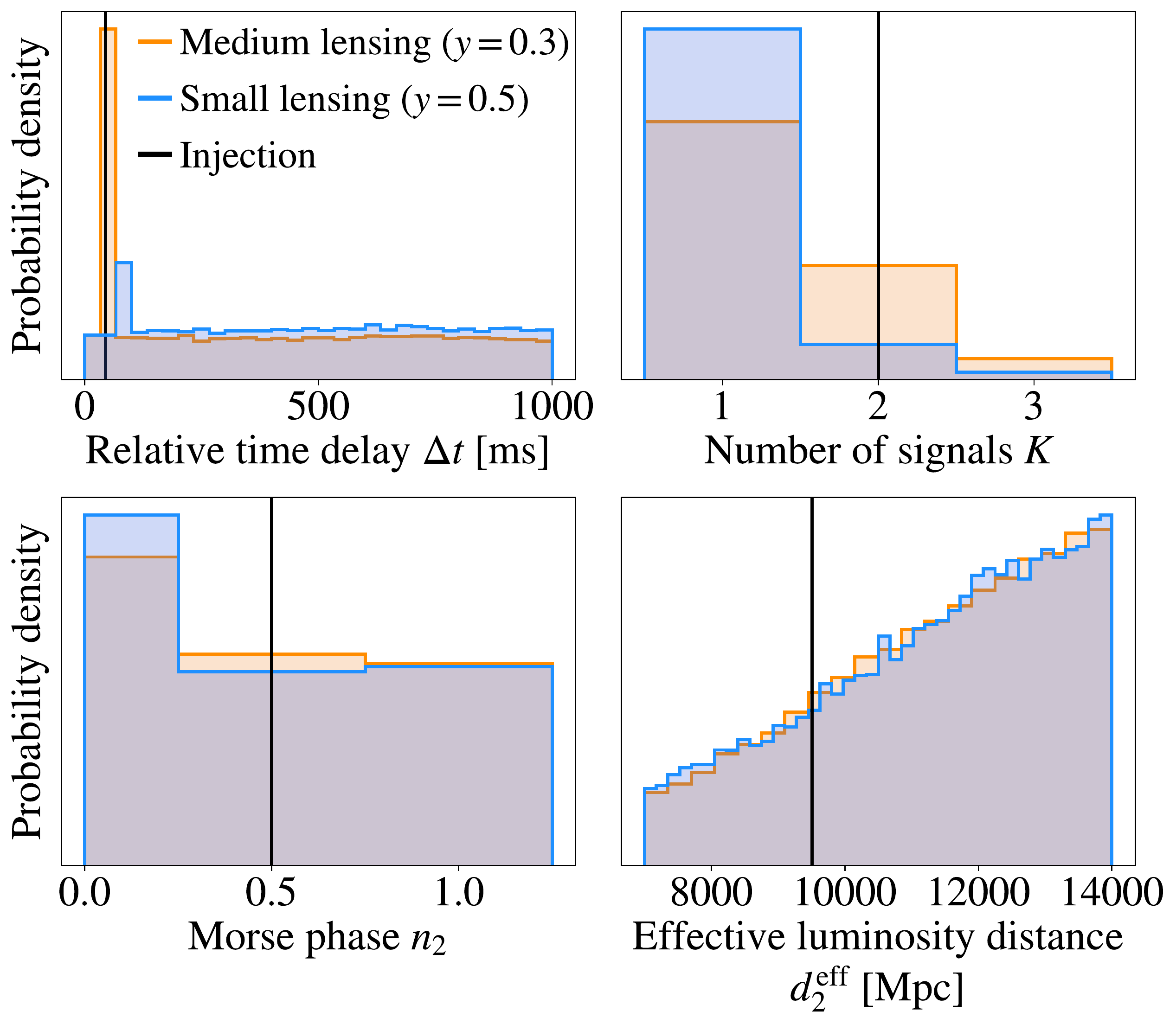}
    \caption{Recovery of lensing parameters in the presence of precession ($\chi_p=0.64$) at SNR 8 for small and medium-lensing cases ($y=0.5$ and $y=0.3$) for a high mass system $(30, 35) M_\odot$. 
    The posteriors obtained do not recover injected values (vertical black lines) correctly, even for medium-lensing case, suggesting that it is not possible to recognize the lensing effect at the SNR of 8.}
    \label{fig:snr8_lens}
\end{figure}
\begin{figure}[htp]
    \centering
    \includegraphics[width=\columnwidth]{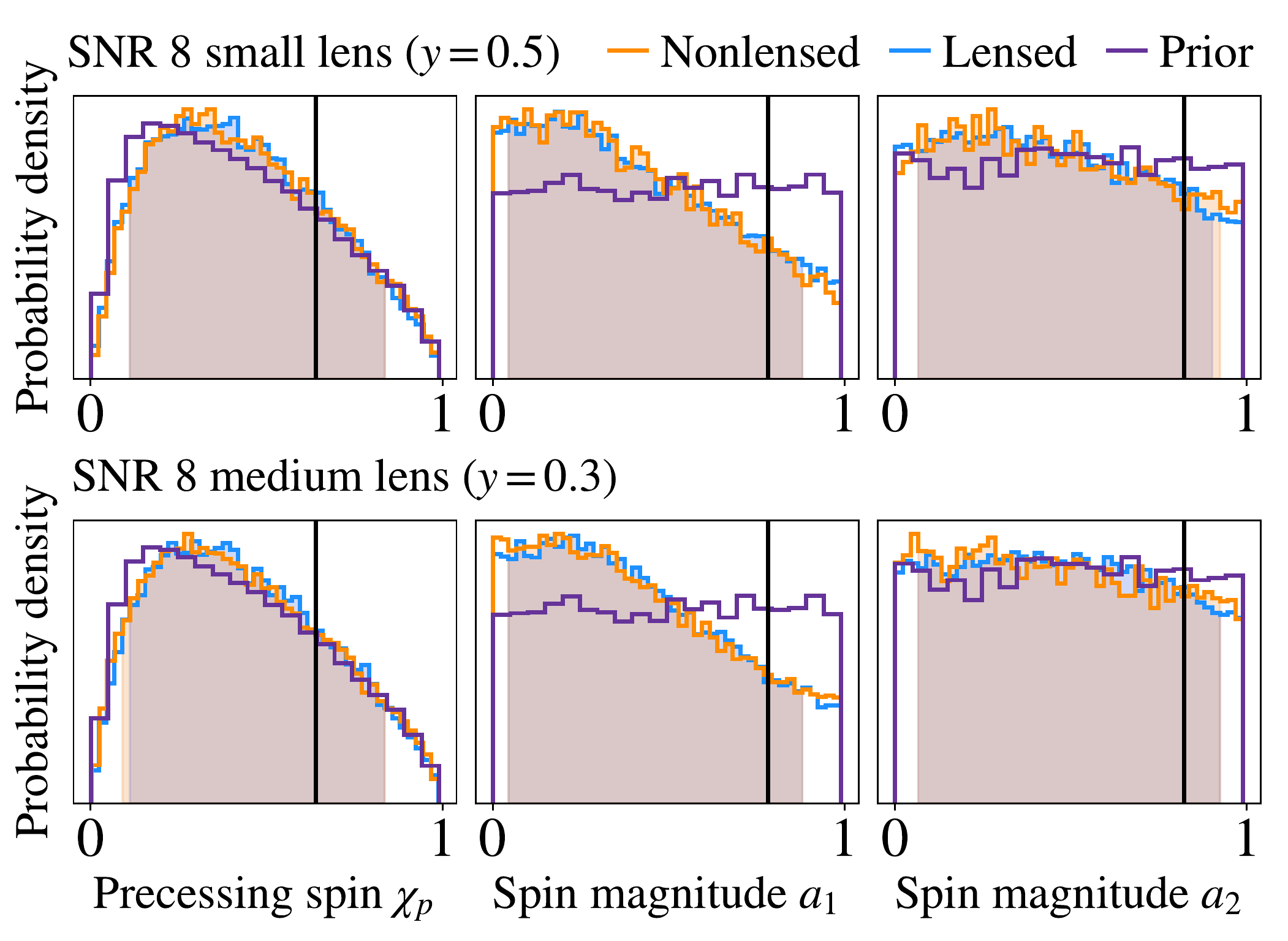}
    \caption{Spin parameters recovered from high mass $(30, 35) M_\odot$ high spin $\chi_p = 0.64$ injections with SNR 8 and varying lensing strength. Shaded regions correspond to 90\% C.I. No significant difference was found in the inclusion of lensing in the recovery model.
    Both models do not lead to informative posteriors for both small and medium-lensed case, suggesting the SNR of 8 is not sufficient to correctly infer precession.}
    \label{fig:snr8_spins}
\end{figure}

We also performed lensed and precessing GW injections with an SNR of 8, which corresponds to the detection threshold SNR. 
We injected two signals with small ($y=0.5$) and medium ($y=0.3$) lensing effects and high spin ($\chi_p = 0.64$).
We recovered the two signals using the PL and PU models. 
From the PL recovery, we find that the number of millilensed GWs is not recovered correctly for both small- and medium-lensing cases, with the posterior suggesting $K=1$ component millilensed GW (Fig.~\ref{fig:snr8_lens}).
The obtained posteriors for the remaining lensing parameters ($\Delta t, n_2, d_2^{\rm{eff}}$) are not informative.
For the recovery of precession, we compared the lensed and unlensed recovery of spin components and effective precessing spin in Fig.~\ref{fig:snr8_spins}.
The lensed and unlensed recoveries follow the same distributions for all three parameters, not leading to informative posteriors.
To check for sampler convergence of this analysis, we tested the recovery with an increasing number of sampler live points which led to the same results, suggesting that an SNR of 8 is not sufficient to recover the precession or lensing parameters.

Our results confirm that in the low $\text{SNR}=8$ regime, it becomes exceedingly difficult to recover lensing or precession from the data, as expected. 
Moreover, the correct recovery of the number of millilensed GWs is hindered, thereby highlighting the inadequate capture of lensing effects.  
Considering the exploratory nature of this study, our conclusions regarding differentiating precession and lensing are, therefore, limited to and based on studies with injected SNR of 18 and higher.

\section{Conclusions}
\label{sec:conclusions}

\begin{table*}[]
    \centering
    \caption{Summary results of injections with \textbf{varying spin} for low masses (14, 10) $M_\odot$, medium-lensing $y=0.3$, and $\textrm{SNR}=18$. All injections are with lensed precessing model, the recovery models are: PL - precessing lensed, PU - precessing unlensed, NL - nonprecessing lensed. We find that we can correctly infer lensing information for all spinning cases, both with and without accounting for precession. The recovery of precession is inaccurate for medium and high spins, suggesting the SNR of 18 is insufficient to infer precession information from a lensed and moderately/highly precessing GW.}
    \begin{tabularx}{1.\linewidth}{c @{\extracolsep{\fill}} c @{\extracolsep{\fill}} c @{\extracolsep{\fill}} c}
        \toprule
        Spin $\chi_p$  &  Recovery model & Number of lensed signals recovered correctly & Is precession recovered correctly? \\
        \hline
        \multirow{3}{*}{\textbf{0.16}} & PL & Yes, see Fig.~\ref{fig:dL_dt_small_mass_vary_spin} & Yes, see Fig.~\ref{fig:a1_a2_slow_high_spin} \\
        & NL & Yes, see Fig.~\ref{fig:deff2_dt_prec_vs_nonprec} & -  \\ 
        & PU & -  & Yes, see Fig.~\ref{fig:a1_a2_slow_high_spin}\\
        \hline 
        \multirow{3}{*}{\textbf{0.36}} & PL & Yes, see  Fig.~\ref{fig:dL_dt_small_mass_vary_spin} & No, see Fig.~\ref{fig:a1_a2_slow_high_spin} \\
        & NL & Yes, see Fig.~\ref{fig:deff2_dt_prec_vs_nonprec} & - \\
        & PU & - & No, see Fig.~\ref{fig:a1_a2_slow_high_spin} \\
        \hline 
        \multirow{3}{*}{\textbf{0.64}} & PL & Yes, see Fig.~\ref{fig:dL_dt_small_mass_vary_spin}  & No, see Fig.~\ref{fig:a1_a2_slow_high_spin}\\ 
        & NL & Yes, see Fig.~\ref{fig:deff2_dt_prec_vs_nonprec} & - \\
        & PU & - & No, see Fig.~\ref{fig:a1_a2_slow_high_spin}\\
        \hline
        \hline
    \end{tabularx}
    \label{tab:summary_vary_spin}
    \hspace{1em}
        \caption{Summary results of \textbf{vary lensing} for higher masses (30, 35) $M_\odot$, high spin $\chi_p=0.64$, and $\textrm{SNR}=18$ injections.
        Compared to the lower mass case from Table~\ref{tab:summary_vary_spin}, for higher BH masses we find more bias present in the recovered lensing parameters (in particular $d_2^\text{eff}$). The precession information is also largely uninformative.
        }
    \begin{tabularx}{1.\linewidth}{c @{\extracolsep{\fill}} c @{\extracolsep{\fill}} c @{\extracolsep{\fill}} c}
        \toprule
         Lensing & Recovery model &  
         Number of lensed signals recovered correctly & Is precession recovered correctly? \\
         \hline
         \multirow{2}{*}{\textbf{small ($\mathbf{y\!=\!0.5}$)}} & PL & Yes, see Fig.~\ref{fig:snr18_large_vs_small_lens} & \multirow{6}{*}{Non-informative posterior distributions}\\
         & PU & - & \multirow{6}{*}{for spin magnitudes, see Fig.~\ref{fig:component_spins_vary_lens}} \\
         \cline{1-3}
         \multirow{2}{*}{\textbf{medium ($\mathbf{y\!=\!0.3}$)}} & PL & Yes, see Fig.~\ref{fig:snr18_snr40_comparison} & \\ 
         & PU & - & \\
         \cline{1-3}
         \multirow{2}{*}{\textbf{large ($\mathbf{y\!=\!0.1}$)}} & PL & Yes, see Fig.~\ref{fig:snr18_large_vs_small_lens} & \\
         & PU & - & \\ 
         \hline
         \hline
    \end{tabularx}
    \label{tab:summary_vary_lens}
    \hspace{1em}
    \caption{Summary results of \textbf{varying SNR} for higher masses (30, 35) $M_\odot$, and high spin $\chi_p = 0.64$ injections. We find that the detector-threshold SNR 8 is not sufficient to recover and distinguish precession and millilensing. For the SNR of 18 cases we can recover lensing information in the presence of precession, but find biases in the spin recoveries. The SNR of 40 is sufficiently high to recover both lensing and precession, however, ignoring lensing can lead to biases in spin recoveries.}
    \begin{tabularx}{1.\linewidth}{c @{\extracolsep{\fill}} c @{\extracolsep{\fill}} c @{\extracolsep{\fill}} c @{\extracolsep{\fill}} c}
        \toprule
        SNR & Lensing & Recov. model & Lensed signals recovered correctly & Is precession recovered correctly? \\
        \hline
        \multirow{4}{*}{\textbf{8}} & \multirow{2}{*}{small ($y=0.5$)} & PL & No, see Fig.~\ref{fig:snr8_lens}  & No, see Fig.~\ref{fig:snr8_spins} \\
        & & PU & -  & No, see Fig.~\ref{fig:snr8_spins} \\ 
        \cline{2-5}
        & \multirow{2}{*}{medium ($y=0.3$)} & PL & No, see Fig.~\ref{fig:snr8_lens} & No, see Fig.~\ref{fig:snr8_spins} \\
        & & PU & - & No, see Fig.~\ref{fig:snr8_spins}\\ 
        \hline 
        \multirow{2}{*}{\textbf{18}} & \multirow{2}{*}{medium ($y=0.3$)} & PL & Yes, see Fig.~\ref{fig:snr18_snr40_comparison} & $a_1, a_2$ noninformative, $\chi_p$ agrees with injected value, see Fig.~\ref{fig:snr18_snr14_spins} \\
        & & PU & - & Spins biased toward lower values, see Fig.~\ref{fig:snr18_snr14_spins} \\
        \hline
        \multirow{2}{*}{\textbf{40}} & \multirow{2}{*}{medium ($y=0.3$)} & PL & Yes, see Fig.~\ref{fig:snr18_snr40_comparison}  & Yes, see Fig.~\ref{fig:snr18_snr14_spins}\\
        & & PU & -  & Spins biased toward extreme values, see Fig.~\ref{fig:snr18_snr14_spins} \\
        \hline
        \hline
    \end{tabularx}
    \label{tab:summary_vary_SNR}
\end{table*}

\begin{figure*}[t!]
    \includegraphics[width=1.\linewidth]{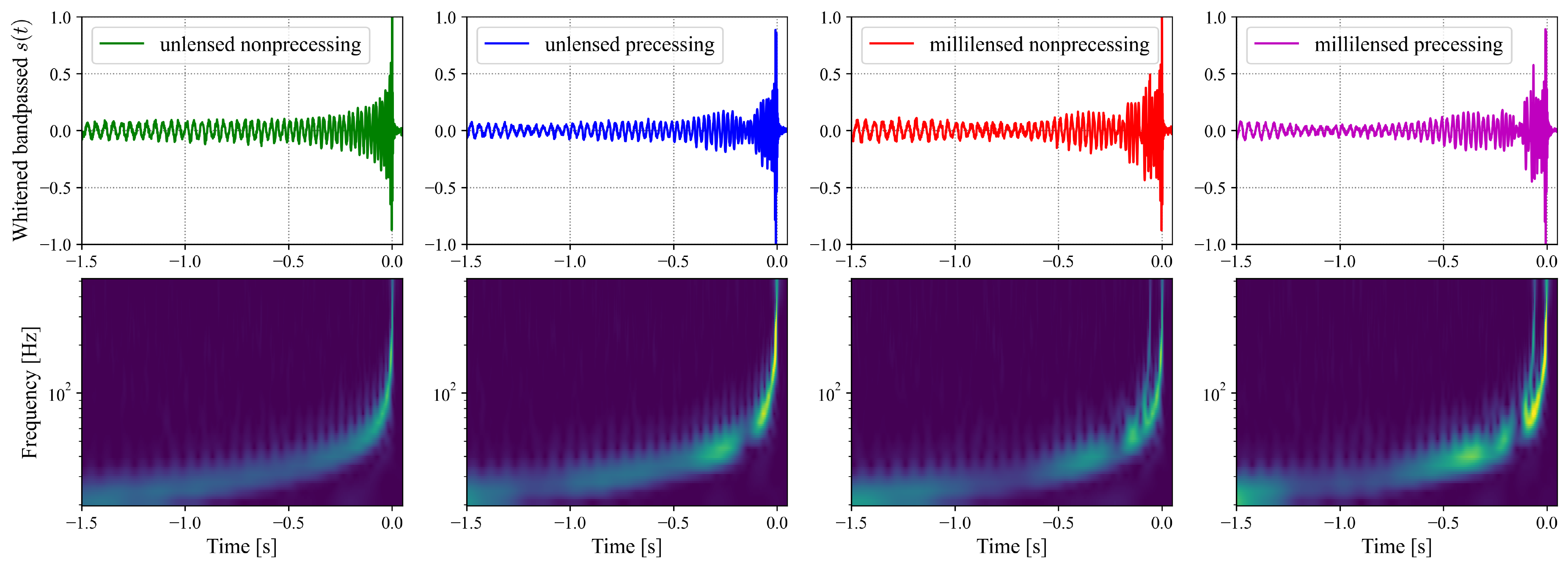} 
\caption{The same signals shown in Fig.~\ref{fig:wf_noise_free} but injected into the noise data of the Einstein Telescope's design PSD. One can see that the distinguishability between unlensed precessing and millilensed nonprecessing signals is significantly easier for the ET compared to the same signals in LVK detectors, presented in Fig.~\ref{fig:wf_noise_free}. However, it is still nontrivial to identify the origin of the millilensed precessing signal from millilensed nonprecessing one. This result addresses the necessity of conducting further analysis such as the PE of this work.}
\label{fig:wf_noisy_et}
\end{figure*}

In this study, we examined the ability to distinguish GW lensing and precession in simulated GW signals detectable by LIGO-Virgo detectors. 
From an example millilensed, precessing BBH signal of 30$M_\odot$--10$M_\odot$ and $\chi_p = 0.76$, we found that the precession effect is not degenerate with millilensing, and we can distinguish the two effects both by comparing SNRs obtained with different templates, as well as correctly recovering lensing and precession parameters via a PE study at aLIGO's design PSD. 
Furthermore, we performed additional PE studies for LIGO O4 sensitivity over a range of injected signals with varying source masses, spins, lensing strengths and SNRs with results summarized in Tables~\ref{tab:summary_vary_spin}-~\ref{tab:summary_vary_SNR}.
In general, we found that lensing is recovered correctly for more cases than precession. 
In particular, the relative time delay between lensed GWs was recovered correctly, both when accounting for or ignoring the precession effect, for the signals with injected SNRs of 18 and above.
The effective luminosity distance, which corresponds to the magnification of the lensed GWs, is found to be sensitive to the precession effect and exhibits biases when neglecting precession in spinning systems with $\chi_p\sim 0.6$ and above. 
The recovery of effective precessing spin $\chi_p$ and dimensionless spin components $a_1$ and $a_2$ from lensed precessing signals was found to be dependent on the injected lensing strength.
In the small- and medium-lensed cases, neglecting the lensing effect did not affect the recovery of BH spins at SNR 18. 
However, for large lensing scenarios ($y=0.1$), the results of lensed and unlensed analyses lead to significant differences in recovered BH component spins.
We conclude that it is possible to recover lensing for signals with SNRs of 18 and above. 
The precession analyses, however, lead to uninformative spin posteriors for signals with SNR 18, subject to biases in the large-lensing scenarios.
This could be improved with SNR, as demonstrated for the SNR 40 case, where accounting for lensing played an important role in the recovery of spins.

Furthermore, when looking at the same example signals in Sec.~\ref{sec:test_examples} injected into the Einstein Telescope's design sensitivity noise model~\cite{Hild:2010id} (Fig.~\ref{fig:wf_noisy_et}), we note that the identification of millilensed and precessing signals becomes much easier. 
Indeed, we can see the whitened bandpassed $s(t)$ are quite similar to the noise-free $h(t)$. Additionally, from the spectrogram, we can observe much clearer patterns than in Fig.~\ref{fig:wf_noisy_aligo}, although distinguishing nonprecessing and precessing signals from the millilensed cases is still quite challenging. 
For these signals, the SNR-based test based on unlensed or millilensed hypotheses along with nonprecessing or precessing template waveforms can be used to obtain initial intuition for the different origins of a given target signal.
For example, if the PL target is given, we can obtain $\textrm{SNR}=411.4$ with the PL template, while we get lower SNRs, $144.0$ and $107.6$, with NU and NL templates, respectively. However, this approach still requires multiple tests with templates of different hypotheses for a single target signal. 
Therefore, if a proper PE-based analysis is performed in the case of ET signals, analogous to the study for  LVK  detectors presented in this work, it will be possible to recognize whether the millilensed event is from a precessing or nonprecessing binary.
However, a detailed investigation of this approach in the context of ET is beyond the scope of this paper, and we leave this for future work.

The findings reported here shed new light on the confusion between lensing and precession effects in LVK-like GW signals, providing a first study of distinguishing precessing and millilensed GWs, to the best of our knowledge.
However, there are several limitations which could be further explored in future work.
In the SNR-based study, we considered GW millilensing by a pointlike lens of $M_{lz}\!=\!10^{3.2}M_\odot$ in the geometrical optics limit. 
It is important to note that extended lens mass distributions in quasigeometrical optics approximation~\cite{Takahashi:2004mc} or more complex lensing configurations in wave optics regime (e.g.,~\cite{Diego:2019lcd,Seo_microlensing}) may introduce more complexity to the observed signals and pose additional challenges in their correct recovery and interpretation.
In the PE study, we applied a phenomenological millilensing framework, which allows oe to account for arbitrary lens configurations.
It is important to note that while this framework is primarily based on the geometric optics approximation, it is applicable to lens masses starting from the order of 1000 $M_\odot$ for LVK frequencies~\cite{Liu2023exploring}, which corresponds to the mass range considered in this paper.
The impact of more complex lensing configurations in the wave optics limit is an important area for future study.

Furthermore, broader studies such as a systematic study done in~\cite{Bondarescu:2022srx} could provide more comprehensive findings relating the effects of gravitational lensing and spin-induced precession. 
In particular, finding a decisive criterion for the discrimination between the two effects would be helpful for future observations on GWs containing various intrinsic and extrinsic characteristics. 

While our study was limited to lensed precessing signals, other physical effects, including eccentricity, can also introduce modulation in GW waveforms~\cite{LIGOScientific:2019dag, Romero-Shaw:2020thy, Romero-Shaw:2021ual, Romero-Shaw:2022xko}.
Detailed studies of identifying the origin of waveform modulation and quantifying the degeneracies between lensing, precession and eccentricity would be a fruitful area for further work, which is beyond the scope of this paper.  

While the waveform approximants employed in this study incorporate higher multipoles, further exploration of their specific impact on the GW analysis could yield valuable insights.
Higher multipoles in GWs can help to overcome the inclination-luminosity distance degeneracy, hence allowing for better spin measurements and constraints of precession which could also improve the ability to distinguish precession from lensing~\cite{fairhurst2020will, fairhurst2020two}.
Furthermore, an analytical study, such as the Fisher-information matrix formalism, or computation of mismatch between lensed and precessing waveforms could provide a complementary answer to the distinguishability of gravitational lensing and spin precession.

\acknowledgements
The authors thank Otto A. Hannuksela, Tjonnie G. F. Li, and Eungwang Seo for fruitful discussions and comments. This work is supported by the National Research Foundation of Korea (NRF) grant funded by the Ministry of Science and ICT (MSIT) of the Korea Government (NRF-2020R1C1C1005863). The work is also partially supported by the Croucher Innovation Award from the Croucher Foundation Hong Kong. The work of K.K. is partially supported by the Korea Astronomy and Space Science Institute (KASI) under the R\&D program (Project No. 2024-1-810-02) supervised by the MSIT. K.K. also acknowledges that the computational work reported in this paper was partly performed on the KASI Science Cloud platform supported by KASI. 
A.L. acknowledges support by grants from the Research Grants Council of Hong Kong (Project No. CUHK 14304622 and 14307923), the start-up grant from the Chinese University of Hong Kong, and the Direct Grant for Research from the Research Committee of The Chinese University of Hong Kong.

\appendix
 
\section{Source parameters recovery with and without accounting for precession and millilensing}

\begin{figure}[t!]
    \centering
    \includegraphics[width=\columnwidth]{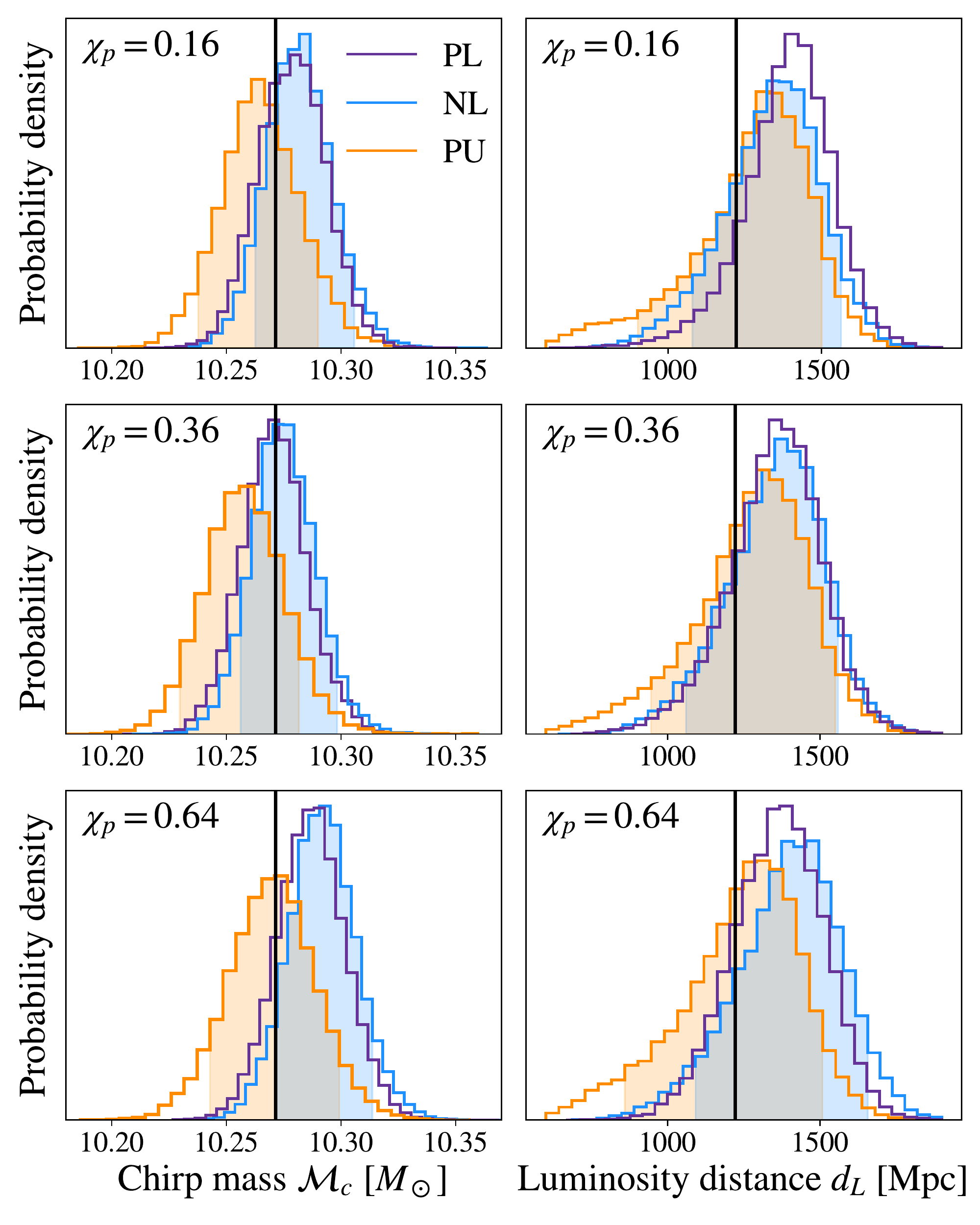}
    \caption{Posterior distributions for the chirp mass $\mathcal{M}_c$ (left column) and luminosity distance $d_L$ (right column), injected with (14,10) $M_\odot$ source masses, lensed and precessing GW model and recovered with precessing and lensed (PL), lensed-only (L) and precessing-only (P) models, for three values of spins: $\chi_p =(0.16, 0.36, 0.64)$ with SNR of 18. The vertical black lines represent the injected values. The shaded regions correspond to 90\% C.I. of the precessing-only (orange) and lensed-only (blue) distributions.
    We note that the PL and L models follow similar distributions, with a preference toward higher values than the P-only recovery. 
    The difference between the three distributions is largest for the $\chi_p=0.64$ case, when the contribution from precession is largest. 
    While this result provides insights into how chirp mass recovery changes across different models, it is not a key finding and serves primarily as a check.
    It does not suggest significant differences that could be used as an indication of the presence of precession or lensing, as any bias falls within 90\% CI. 
    Additionally, the bias is affected by the choice of priors, and could also arise from correlations between parameters, such as the distance-inclination degeneracy which is further modulated by the recovery with precession. 
    This complex situation is beyond the scope of this work and warrants further investigation.
    }
    \label{fig:mchirp_dL_combined}
\end{figure}

We compare the recovery of BBH source parameters, considering the effects of precession and gravitational lensing individually, as well as their combined influence on the recovered parameters.
In Fig.~\ref{fig:mchirp_dL_combined} we present the results obtained for the chirp mass $\mathcal{M}_c$ and the luminosity distance $d_L$ posteriors, for 3 different injected effective precessing spin values $\chi_p =(0.16, 0.36, 0.64)$, recovered using the PL, PU and NL models.
We aim to assess the impact of neglecting lensing or precession on the recovery of chirp mass, which is the parameter best recovered from GW PE.

We show the results for the lower total-mass $(14, 10) M_\odot$ system here, as both precession and gravitational lensing effects accumulate with the inspiral.
Hence, any bias due to the two effects is expected to be more prominent for longer-duration (lower-mass) signals.
As observed in the figures, the injected value remains within the shaded region corresponding to 90\% credible interval (C.I.) of the recovered posterior distributions for chirp mass recovery, however its position w.r.t. the peak of each posterior distribution changes for 3 different spin cases.   
Overall, the precession-only model (PU) tends to prefer lower values of chirp mass for all three spin scenarios, as compared to the lensing-only (NL) and precessing-lensed (PL) models.
The PL and NL models show very similar distributions, with their largest difference occurring for the highest spin $\chi_p = 0.64$ case.
The NL distribution also shows the largest bias in chirp mass for the highly spinning case, where the injected value is at the edge of its 90\% CI.
This suggests that highly-spinning systems ($\chi_p\approx0.6$ and higher) are subject to most biases arising from precession and millilensing.

The luminosity distance recovery in all cases shows a preference for larger values as compared to the injection. 
This is likely due to the power-law prior distribution with spectral index $\alpha =2$ used for $d_L$ which may cause the distribution to prefer higher values.
Similar to chirp mass, the PU recovery of luminosity distance $d_L$ shows a preference for smaller values as compared to the other two models.
Again, the posterior distributions obtained with NL and PL recoveries are very similar to each other, recovering the same range of posterior values.
The high-spin $\chi_p =0.64$ recovery demonstrates the largest bias between the three recovered distributions. 

The fact that the PU model results showed the largest bias compared to the PL recovery in Fig.~\ref{fig:mchirp_dL_combined}, especially prominent in the chirp mass posterior distributions, is further supported by the BFs obtained for the three models, which show that precession-only recovery is disfavored for all three spin values, as listed in Table~\ref{tab:BFs}.
Nevertheless, the bias studied for these analyses does not exceed the 90\% C.I. for the injected SNR of 18.
Hence, we expect that it is still possible to identify the presence of precession in the signals.

\section{Prior validation}
\label{sec:appendix_priors}

\begin{figure}[h!]
    \centering
    \includegraphics[width=0.9\linewidth]{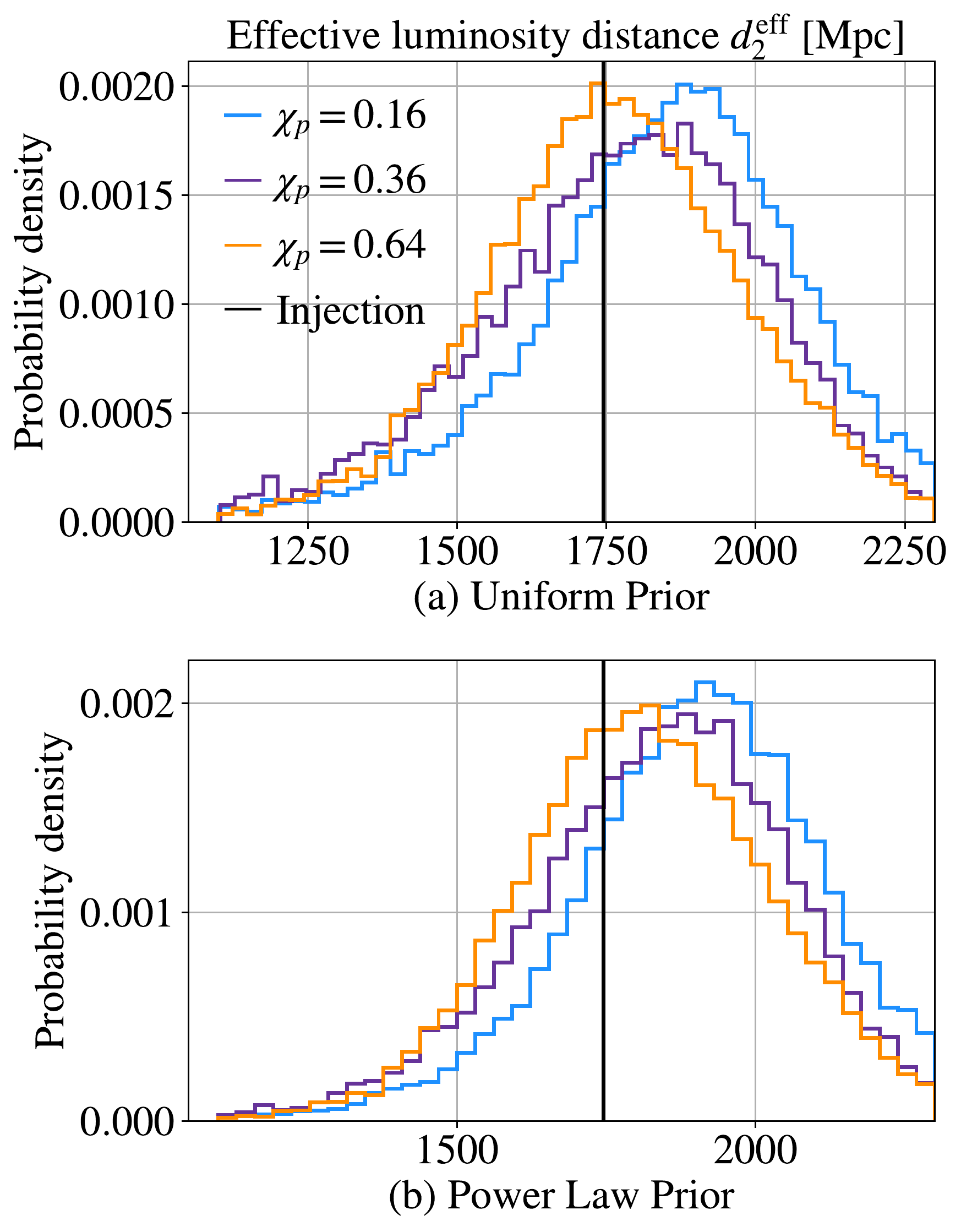}
    \caption{Recovery of the effective luminosity distance $d_2^\mathrm{eff}$ from varying-spin injections, comparison of (a) uniform and (b) power-law prior distributions.}
    \label{fig:dL2_prior_comparison}
\end{figure}

In Fig.~\ref{fig:dL2_prior_comparison} we show a comparison of recovered effective luminosity distance $d_2^\mathrm{eff}$ distributions using uniform and power-law prior distributions to validate the trends of distributions shifting toward larger values for smaller injected spins. 
As can be noted, while the uniform-prior results are closer distributed to the injection value, the trend remains the same in both plots. 
It suggests that this trend is not a direct consequence of the prior choice used for $d_2^\mathrm{eff}$.

\bibliography{references}

\end{document}